\RequirePackage[l2tabu, orthodox]{nag}
\documentclass[journal, 10pt, twocolumn, letterpaper, oneside]{IEEEtran}

\IEEEoverridecommandlockouts

\usepackage{ifpdf}

\usepackage[noadjust]{cite}

\ifCLASSINFOpdf
  \usepackage[pdftex]{graphicx}
\else
  \usepackage[dvips]{graphicx}
\fi

\usepackage[cmex10]{amsmath}
\usepackage{array}

\usepackage{fixltx2e}

\usepackage{url}
\usepackage[utf8]{inputenc} 
\usepackage[T1]{fontenc}
\usepackage{ifthen}

\usepackage{mathtools}
\usepackage{amssymb}
\usepackage{relsize}
\usepackage{bbm}
\usepackage{xfrac}
\usepackage{amsthm}
\theoremstyle{plain}
\newtheorem{definition}{Definition}
\newtheorem{lemma}{Lemma}

\newtheorem{theorem}{Theorem}

\newtheorem{remark}{Remark}

\usepackage[varg, cmbraces]{newtxmath}

\newcommand*{\mydprime}{^{\prime\prime}\mkern-1.2mu}
\newcommand*{\mytrprime}{^{\prime\prime\prime}\mkern-1.2mu}

\usepackage{accents}

\usepackage{xcolor}
\definecolor{burgundy}{rgb}{0.545098,0,0}
\definecolor{navyblue}{rgb}{0.0, 0.0, 0.5}
\definecolor{leafgreen}{rgb}{0.290196, 0.470588, 0.0}
\definecolor{bluegreen}{rgb}{0, 0.470588, 0.415686}
\definecolor{zuhl}{rgb}{0.1875, 0.26171875, 0.46484375}
\definecolor{orange}{rgb}{1, 0.6470588235, 0}
\definecolor{red}{rgb}{1, 0, 0}

\usepackage{color}
\usepackage{overpic}
\usepackage{pict2e}
\usepackage{booktabs}

\usepackage{tikz}
\usepackage{pgfplots}

\newcommand{\lemref}[1]{Lemma~\ref{#1}}

\newcommand{\thref}[1]{Theorem~\ref{#1}}
\newcommand{\defref}[1]{Definition~\ref{#1}}

\newcommand{\sectref}[1]{Section~\ref{#1}}

\newcommand{\appref}[1]{Appendix~\ref{#1}}

\newcommand{\cX}{S(X)}
\newcommand{\cZ}{S(Z)}

\usepackage{soul}
\setstcolor{red}

\allowdisplaybreaks[4]

\usepackage{array}
\usepackage[linesnumbered, algoruled, boxed, lined, vlined]{algorithm2e}
\SetKwRepeat{Do}{do}{while}

\usepackage{hyperref}
\hypersetup{
    colorlinks,
    linkcolor = {red!60!black},
    citecolor = {blue!60!black},
    urlcolor = {blue!50!black}
}

\usepackage{microtype}

\begin{document}

\title{Third-Order Asymptotics of   Variable-Length Compression Allowing Errors}

\IEEEoverridecommandlockouts

\author{%
\IEEEauthorblockN{%
Yuta~Sakai,~\IEEEmembership{Member,~IEEE,} Recep~Can Yavas,~\IEEEmembership{Graduate Student Member,~IEEE,} and
Vincent~Y.~F.~Tan,~\IEEEmembership{Senior~Member,~IEEE,}}%
\thanks{This work is supported by a Singapore National Research Foundation (NRF) Fellowship (R-263-000-D02-281). This work was presented in part at the International Symposium on Information Theory and Its Applications (ISITA) in Kapolei, Hawaii, USA in October 2020.}
\thanks{Y.~Sakai is with the Department of Electronics and Computer Science,
Graduate School of Engineering,
University of Hyogo, Email: \url{yuta.sakai@eng.u-hyogo.ac.jp}. R.~C.~Yavas is with the Department of Electrical Engineering, California Institute of Technology, Email: \url{ryavas@caltech.edu}. 
V.~Y.~F.~Tan is with the Department of Electrical and Computer Engineering and the Department of Mathematics, National University of Singapore, Singapore, Email: \url{vtan@nus.edu.sg}.}%
}%

\maketitle

\begin{abstract}
This study investigates the fundamental limits of variable-length compression in which prefix-free constraints are not imposed (i.e., one-to-one codes are studied) and non-vanishing error probabilities are permitted. 
Due in part to a crucial relation between the variable-length and fixed-length compression problems, our analysis requires a careful and refined analysis of the fundamental limits of fixed-length compression in the setting where the error probabilities are allowed to approach either zero or one polynomially in the blocklength.
To obtain the refinements, we employ tools from moderate deviations and strong large deviations. 
Finally, we provide the third-order asymptotics for the problem of variable-length compression with non-vanishing error probabilities.
We show that unlike several other information-theoretic problems in which the third-order asymptotics are known, for the problem of interest here, the third-order term depends on the permissible error probability.
\end{abstract}


\IEEEpeerreviewmaketitle

\section{Introduction}

Characterizing fundamental limits of  coding problems is the central goal in information theory.
The class of variable-length compression problems  (i.e., fixed-to-variable length coding problems) constitute a  classical and important family of information-theoretic problems in view of their multitude of practical applications.
Han \cite{han_2000} considered the problem of variable-length compression with prefix-free constraints allowing a \emph{small (i.e., vanishing) error probability.}
He then derived the first-order optimal coding rate for a general source when the error probability is required to vanish with increasing blocklengths.
Later, Koga and Yamamoto \cite{koga_yamamoto_2005} derived the first-order optimal coding rate in the regime of non-vanishing error probabilities.
In the particular case of a stationary memoryless source $X$, their work \cite{koga_yamamoto_2005} showed that
\begin{align}
L_{\mathrm{prefix}}^{\ast}(\varepsilon \mid X^{n})
=
n \, (1 - \varepsilon) \, H(X) + \mathrm{o}( n ) \label{eqn:L_prefix}
\end{align}
as $n \to \infty$ for fixed $0 \le \varepsilon \le 1$, where $L_{\mathrm{prefix}}(\varepsilon \mid X^{n})$ denotes the minimum of average codeword lengths of binary prefix-free codes for $n$ i.i.d.\ copies $X^{n}$ of $X$ in which the error probability is at most $\varepsilon$, and $H(X)$ stands for the entropy of $X$ measured in bits. Hence, in general, the strong converse property (cf.~\cite{wolfowitz_1978}) fails to hold in variable-length compression problems. The equality in~\eqref{eqn:L_prefix} also demonstrates the utility in using both {\em variable-length coding} as well as permitting {\em non-vanishing error probabilities}. This is because the first-order fundamental limit is reduced to $(1-\varepsilon)\, H(X)$  if we permit an error probability $\varepsilon > 0$. If one were to use fixed-length codes or demand that the error probabilities are vanishing, one cannot compress a source $X$ with entropy $H(X)$ with rate strictly below $H(X)$. This formalism is an alternative to lossy source coding (rate-distortion) for the purpose of reducing the  compression rate. In  lossy compression, the source suffers from some distortion; here the source sequence is  either reproduced perfectly or an arbitrary sequence is generated (with probability~$\varepsilon$). %This is somewhat  reminiscent of the practically-relevant erasure decoding setting, in which the code never outputs an incorrect source sequence, but
%it is permitted to output a special ``don't know'' symbol with a certain probability. In our case, the ``erasure probability'', i.e., %the probability that the decoder outputs an arbitrary sequence,  is~$\varepsilon$. 

In this paper, we consider variable-length compression problems \emph{without} prefix-free constraints.
In the zero-error setting, this class of fixed-to-variable length codes is often known as \emph{one-to-one codes.}
While the redundancy%
\footnote{The redundancy of a fixed-to-variable length code is defined as ``its average codeword length minus the entropy of the source.''}
of a prefix-free code is always nonnegative, the redundancy of a one-to-one code can be \emph{negative} (cf.\ \cite{wyner_1972, alon_orlitsky_1994}).%
\footnote{Hence, the redundancy of a one-to-one code is sometimes termed as the \emph{anti-redundancy} (cf.\ \cite{szpankowski_2008}).}
In fact, Szpankowski and Verd\'{u} \cite{szpankowski_verdu_2011} proved an asymptotic expansion of the smallest redundancies of one-to-one codes for a stationary memoryless source $X$.
They showed that for finitely supported non-equiprobable $X$,
\begin{align}
L^{\ast}(0 \mid X^{n})
=
n \, H(X) - \frac{ 1 }{ 2 } \log n + \mathrm{O}( 1 )
\label{eq:SV_3rd}
\end{align}
as $n \to \infty$, where $L^{\ast}(0 \mid X^{n})$ denotes  the minimum of average codeword lengths of one-to-one codes for $X^{n}$.
Furthermore, Szpankowski \cite{szpankowski_2008} refined the remainder term $+\mathrm{O}( 1 )$ in \eqref{eq:SV_3rd} when $X$ is a Bernoulli source, and clarified necessary and sufficient conditions on $X$ for which the dominant term within the  $+\mathrm{O}( 1 )$ remainder term converges or oscillates.
On the other hand, in the regime of non-vanishing error probabilities, Kostina, Polyanskiy, and Verd\'{u} \cite{kostina_polyanskiy_verdu_2015} derived the second-order optimal coding rate of this fundamental limit for a stationary memoryless source.
They \cite{kostina_polyanskiy_verdu_2015} showed that
\begin{align}
L^{\ast}(\varepsilon \mid X^{n})
=
n  (1 \!- \!\varepsilon)  H(X) \!-\! \sqrt{ \frac{ n \, V(X) }{ 2 \pi } }  \mathrm{e}^{-\Phi^{-1}(\varepsilon)^{2} / 2} \!+\! \mathrm{O}( \log n )
\label{eq:KPV_2nd}
\end{align}
as $n \to \infty$ for fixed $0 \le \varepsilon \le 1$, provided that the variance and the absolute  third central moment of the information density $- \log P_{X}(X)$ are positive and finite, respectively, where $L^{\ast}(\varepsilon \mid X^{n})$ stands for the minimum of average codeword lengths of non-prefix-free codes for $X^{n}$ in which the error probability is at most $\varepsilon$, the quantity $V(X)$ stands for the \emph{varentropy} of $X$ measured in bits squared per source symbol (cf.\ \cite{kontoyiannis_verdu_2014}), and $\Phi^{-1}( \cdot )$ stands for the inverse of the Gaussian cumulative distribution function. The intuition for the somewhat unusual dispersion term in \eqref{eq:KPV_2nd} can be found in \cite[Section~II.E]{kostina_polyanskiy_verdu_2015}.
It is clear that \eqref{eq:KPV_2nd} is consistent with \eqref{eq:SV_3rd} when $\varepsilon = 0$ because $\Phi^{-1}(0) = -\infty$ and so the second-order term vanishes when~$\varepsilon=0$.

\subsection{Contributions of This Study}

In this study, we consider   refinements of \eqref{eq:SV_3rd} and \eqref{eq:KPV_2nd} simultaneously. In particular, we generalize Szpankowski and Verd\'{u}'s work \cite{szpankowski_verdu_2011} from the zero-error setting (i.e., $\varepsilon = 0$) to the  setting with non-vanishing error probabilities (i.e., $\varepsilon > 0$). More importantly, we refine the $+\mathrm{O}( \log n )$ remainder term  in Kostina \emph{et~al.}'s second-order asymptotic result \cite{kostina_polyanskiy_verdu_2015}. We show that this term equals $-((1-\varepsilon) \log n)/2$.
To do so, we use a  crucial relation between variable-length and fixed-length codes and their fundamental limits (cf.\ \cite{verdu_isit2007}).
We derive higher-order asymptotics of the variable-length compression problem by leveraging that of the fixed-length compression problem.
In this strategy, we have to consider the fixed-length compression problem in which error probability   approaches zero or one polynomially in the blocklength.
To deal with these sequences of error probabilities that tend to the boundary of the open interval $(0, 1)$, we apply techniques from \emph{moderate deviations} and \emph{strong large deviations}  (cf.\ \cite{feller_1971, petrov_1975, dembo_zeitouni_1998}).
The resulting higher-order asymptotics of the fixed-length compression problem yields our desired third-order asymptotic expansion of the fundamental limit of the variable-length compression problem.
Somewhat interestingly, unlike several other information-theoretic problems in which the third-order asymptotics are known, for the problem of interest here, the third-order term depends on the permissible error probability $\varepsilon$.
Finally, we believe that the new mathematical results derived here (cf.\ Lemmas~\ref{lem:MD} and \ref{lem:SLD}) may be of independent interest in information theory and beyond.

\subsection{Related Works}

\subsubsection{Higher-Order Asymptotics of Fixed-Length Compression}

In view of the recent developments of the second- and third-order asymptotics of coding problems \cite{strassen_1964, hayashi_2008, hayashi_2009, polyanskiy_poor_verdu_2010, tan_2014, moulin_2017, hayashi_2018}, given a source $X$ with a countable source alphabet $\mathcal{X}$, it is well-known that
\begin{align}
\log M^{\ast}(n, \varepsilon)
=
n \, H(X) - \sqrt{ n \, V(X) } \, \Phi^{-1}( \varepsilon ) - \frac{ 1 }{ 2 } \log n + \mathrm{O}( 1 )
\label{eq:third_fixed}
\end{align}
as $n \to \infty$ for fixed $0 < \varepsilon < 1$, provided that the variance and the absolute third central  moment of the information density $- \log P_{X}( X )$ are positive and finite, respectively, where $\log$ denotes the logarithm to the base-$2$ and $M^{\ast}(n, \varepsilon)$ stands for the smallest cardinality a set $\mathcal{A} \subset \mathcal{X}^{n}$ in which the $P_{X^{n}}$-probability of $\mathcal{A}$ is at least $1 - \varepsilon$.
 In his seminal work, Strassen \cite{strassen_1964} derived the fourth-order asymptotics of the fixed-length compression with non-vanishing error probabilities. %; we revisit this in this paper. 
%, i.e., for a finitely supported source $X$,%
%\footnote{The $+\log \log \mathrm{e}$ term in the right-hand side of \eqref{eq:Strassen} arises from the change of the base of logarithms of $V(X)$ inside the logarithm function, because the original statement in \cite{strassen_1964} is calculated in terms of the natural logarithm rather than the binary logarithm.}
%\begin{align}
%\log M^{\ast}(n, \varepsilon)
%=
%n \, H(X) - \sqrt{ n \, V(X) } \, \Phi^{-1}( \varepsilon ) - \frac{ 1 }{ 2 } \log \Big( 2 \pi \mathrm{e}^{\Phi^{-1}(\varepsilon)^{2}} \, n \, V(X) \Big) + \log \log \mathrm{e} - \frac{ \sqrt{V(X)} \, S(X) \, (1 - \Phi^{-1}(\varepsilon)^{2}) }{ 6 } + \mathrm{o}( 1 )
%\label{eq:Strassen}
%\end{align}
%as $n \to \infty$ for fixed $0 < \varepsilon < 1$, provided that the information density $- \log P_{X}(X)$ is a nonlattice random variable (r.v.), where $S(X)$ stands for the skewness of   $- \log P_{X}(X)$.
%Equation~\eqref{eq:Strassen} was derived by applying the \emph{Edgeworth expansion} to the information spectrum of the source $X$ \cite{han_2003}.
 Strassen derived this by considering the Edgeworth expansion, a higher-order asymptotic expansion that goes beyond the central limit theorem (cf.\ \cite{feller_1971, petrov_1975}).
In channel coding problems, Moulin \cite{moulin_2017} established certain bounds on the fourth-order optimal coding rate under some regularity conditions on discrete memoryless channels.
Recently, Hayashi \cite{hayashi_2018} investigated the fourth-order asymptotics of various information-theoretic problems.

%\footnote{In the technical parts of \cite{hayashi_2018}, the present authors were not able to verify the correctness of the use of the Edgeworth expansion for \emph{lattice distributions}; see \cite[Chapter~VI.3]{petrov_1975}.}

\subsubsection{Moderate Deviations Analysis}

In information theory, there are two main types of coding theorems that provide refinements to capacity results, theorems concerning \emph{error exponents} and \emph{second-order asymptotics}.
The former evaluates the exponential decay of error probabilities when coding rates are fixed; the latter evaluates the deviations from the first-order fundamental limits (which are typically of order $1/\sqrt{n}$) when error probabilities are fixed.
The \emph{moderate deviations analysis} of coding problems lie in between these two asymptotic regimes. Moderate deviations  examines the interplay between the sub-exponential decay of error probabilities and the deviation from the first-order fundamental limits which are typically of order $\kappa_{n}/  \sqrt{n}$ where the positive sequence $\kappa_{n} = \omega( 1 ) \cap \mathrm{o}( \sqrt{n} )$ as $n \to \infty$. See \cite[Section~I]{altug_wagner_2014} for earlier works on moderate deviations in information theory.
Most notably, in channel coding, Altu\u{g} and Wagner \cite{altug_wagner_2014} investigated the sub-exponential rate of decay of the  error probabilities when the coding rate approaches the capacity {\em slower} than that in the study of the second-order asymptotics \cite{strassen_1964, hayashi_2008, hayashi_2009, polyanskiy_poor_verdu_2010, tan_2014, hayashi_2018}. Some of these techniques will turn out to be useful for the solution of our problem.
%Chubb, Tan, and Tomamichel \cite{chubb_tan_tomamichel_2017} extended the moderate deviations result in classical channel coding to classical communications over quantum channels.

\subsubsection{Exact Asymptotics of Error Probabilities}

The study of strong large deviations \cite[Theorem~3.7.4]{dembo_zeitouni_1998} and \cite[Chapter~VIII.4]{petrov_1975}, or exact asymptotics, is a refinement of the large deviations principle.
While the rate function in the large deviations principle characterizes the exponential decay of the complementary cumulative distribution function of a sum of independent r.v.'s, the theorems in the study of strong large deviations further characterize its sub-exponential decay, and such sub-exponential terms are often referred to as \emph{pre-factors.}
The classical error exponent analysis of channel coding theorems has been refined in the context of the exact asymptotics of the error probability (cf.\ \cite{altug_wagner_2014_SP, altug_wagner_2014_RC, honda_isit2015, honda_isit2018, altug_wagner_2019}). Again, exact asymptotics will play a crucial role in the estimates of some rates and error probabilities in our work.

\subsubsection{Refined Asymptotics of Variable-Length Compression with the Excess Length Constraint}

Instead of the average codeword length, Merhav \cite{merhav_1991} introduced another performance criterion, namely, the \emph{excess length.} Under this setting, one is interested in   finding a threshold under which the complementary cumulative distribution function of the codeword length  evaluated at this threshold  is suitably upper bounded. 
This excess length constraint is closely related to the fixed-length compression problem (cf.\ \cite{kontoyiannis_verdu_2014}).
Third-order asymptotic expansions under the excess length constraint of \emph{type size codes} for {\em universal} variable-length compression were investigated by Kosut and Sankar \cite{kosut_sankar_2017}.
Iri and Kosut \cite{iri_kosut_2019} generalized this work  by considering parametric sources defined by finite-dimensional exponential families. 
%Saito and Matsushima \cite{saito_matsushima_2017} derived a moderate deviations result under the excess length constraints for \emph{Bayes codes} \cite{clarke_barron_1990}.
Finally, Nomura and Yagi \cite{nomura_yagi_2019} established general formulas for the first- and second-order terms of the fundamental limits of this class of problems for a general source.

\iffalse
\subsubsection{Variable-Length Slepian--Wolf Coding}
The class of Slepian--Wolf prefix-free coding problems was studied by Kuzuoka and Watanabe \cite{kuzuoka_watanabe_2015} and by He, Lastras-Monta\~{n}o, Yang, Jagmohan, and Chen \cite{he_montano_yang_jagmohan_chen_2009}.
Specifically, He \emph{et al.}'s result \cite{he_montano_yang_jagmohan_chen_2009} can be thought of as a moderate deviations result, i.e., they evaluated the second-order coding rate of variable-length Slepian--Wolf coding when error probabilities vanish but the rate of decay is not  exponential in the blocklength. 
Applying Bernstein's inequality, Kuzuoka \cite{kuzuoka_isita2012} discovered an alternative proof of the achievability result in He \emph{et al.}'s work \cite{he_montano_yang_jagmohan_chen_2009}.
This alternative proof is more amenable for combining large deviations analysis and  the usual techniques used in information spectrum methods \cite{han_2003}. \fi

\subsection{Paper Organization}

The rest of this paper is organized as follows:
\sectref{sect:pre} introduces basic definitions and notations in this study.
\sectref{sect:variable} revisits previous works summarized in \eqref{eq:SV_3rd} and \eqref{eq:KPV_2nd}, and states our main result as their integration.
\sectref{sect:fourth} proves our main result by presenting several technical lemmas.
\sectref{sect:block} investigates moderate deviations and strong large deviations analyses for the fixed-length compression.
\sectref{sect:conclusion} concludes this study.

\section{Preliminaries}
\label{sect:pre}

\subsection{Random Variables and Discrete Memoryless Sources}
\label{sect:dms}

In this subsection, we introduce basic notions in probability theory, a discrete memoryless source and its information measures.
Let $(\Omega, \mathcal{F}, \mathbb{P})$ the underlying probability space, and $Z$ a real-valued r.v.
Denote by $P_{Z} \coloneqq \mathbb{P} \circ Z$ the probability distribution induced by $Z$.
We say that $Z$ is a \emph{lattice} r.v.\ if it is discrete and there exists a positive constant $d$ such that $\mathcal{D}(Z) \coloneqq \{ z_{1} - z_{2} \mid P_{Z}( z_{1} ) \, P_{Z}( z_{2} ) > 0 \}$ is a subset of $d \mathbb{Z} \coloneqq \{ \dots, -2d, -d, 0, d, 2d, \dots \}$.
Otherwise, we say that $Z$ is a \emph{nonlattice r.v.}
For a lattice r.v.\ $Z$, its \emph{maximal span} is defined by the maximum of positive constants $d$ satisfying $\mathcal{D}(Z) \subset d \mathbb{Z}$.

Given a real-valued r.v.\ $Z$ and a real number $0 < \varepsilon < 1$, define the \emph{$\varepsilon$-cutoff transformation action of $Z$} \cite[Equation~(13)]{kostina_polyanskiy_verdu_2015} by
\begin{align}
\langle Z \rangle_{\varepsilon}
\coloneqq
\begin{cases}
Z
& \mathrm{if} \ Z < \eta ,
\\
B \, Z
& \mathrm{if} \ Z = \eta ,
\\
0
& \mathrm{if} \ Z > \eta ,
\end{cases}
\label{def:cutoff}
\end{align}
where $B$ is the Bernoulli r.v.\ with parameter $1 - \beta$ in which $B$ is independent of $Z$, and two real parameters $\eta \in \mathbb{R}$ and $0 \le \beta < 1$ are chosen so that
\begin{align}
\mathbb{P}\{ Z > \eta \} + \beta \, \mathbb{P}\{ Z = \eta \}
=
\varepsilon .
\end{align}

Consider a countably infinite alphabet $\mathcal{X}$ and an $\mathcal{X}$-valued r.v.\ $X$.
In this study, i.i.d.\ copies $\{ X_{i} \}_{i = 1}^{\infty}$ of $X$ play the role of a discrete memoryless source, and we simply call $X$ the \emph{source.}
A source $X$ is said to be \emph{finitely supported} if the support $\operatorname{supp}(X) \coloneqq \{ x \in \mathcal{X} \mid P_{X}( x ) > 0 \}$ is finite.
We say that $X$ is a \emph{lattice source} if $\log P_{X}( X )$ is a lattice r.v., where $\log$ stands for the logarithm to the base-$2$.
On the other hand, we say that $X$ is a \emph{nonlattice source} if $\log P_{X}(X)$ is a nonlattice r.v.
For a lattice source $X$, denote by $d_{X}$ the maximal span of $\log P_{X}( X )$.
For convenience, we set $d_{X}$ to be zero if $X$ is a nonlattice source.

 Define the {\em Shannon entropy}, the {\em R\'enyi entropy}, the {\em varentropy}, and one-sixth of the {\em skewness} of $X$ as
\begin{align}
H(X)
& \coloneqq
\sum_{x \in \operatorname{supp}(X)} P_{X}( x ) \log \frac{ 1 }{ P_{X}( x ) } ,
\\
H_{\alpha}(X)
& \coloneqq
\frac{ 1 }{ 1 - \alpha } \log \left( \sum_{x \in \operatorname{supp}(X)} P_{X}( x )^{\alpha} \right),
\\
V(X)
& \coloneqq
\sum_{x \in \operatorname{supp}(X)} P_{X}( x ) \left( \log \frac{ 1 }{ P_{X}( x ) } - H(X) \right)^{2} ,
\\
S(X)
& \coloneqq \frac{1}{6}
\sum_{x \in \operatorname{supp}(X)} P_{X}( x ) \left( \frac{ - \log P_{X}( x ) - H(X) }{ \sqrt{ V(X) } } \right)^{3} ,\label{eqn:skewness}
\end{align}
respectively. 
Throughout this study, assume that $V(X) > 0$, i.e., assume that $X$ is not uniformly distributed on a finite subalphabet $\mathcal{A} \subset \mathcal{X}$.

Similar to a notion in probability theory (cf.\ \cite[Chapter~VIII.2]{petrov_1975}), we define the following condition on a source~$X$.

\begin{definition}
\label{def:Cramer}
We say that a source $X$ satisfies \emph{Cram\'{e}r's condition} if $H_{\alpha}(X)$ is finite for some $0 < \alpha < 1$.
\end{definition}

\begin{remark}
The R\'{e}nyi entropy $H_{\alpha}( X )$ can be thought of as a monotone function of the cumulant generating function of the information density $-\log P_{X}(X)$, i.e., we readily see that
\begin{align}
(1 - \alpha) \, H_{\alpha}( X )
=
\log \mathbb{E}[ 2^{(1 - \alpha) \log P_{X}(X)} ] .
\end{align}
Namely, Cram\'{e}r's condition on $X$ ensures the existence of the $k$-th moment $\mathbb{E}[ \log^{k} P_{X}(X) ]$ for every $k \ge 1$, i.e., the quantities $H(X)$, $V(X)$, and $S(X)$ are finite in this case.
Note that there exists a source $X$ such that $H(X)$, $V(X)$, and $S(X)$ are finite but Cram\'{e}r's condition fails to hold (see, e.g., \cite[Example~5]{kovacevic_stanojevic_senk_2013}).
On the other hand, since $H_{\alpha}(X) \le \log |\operatorname{supp}(X)|$ for every $\alpha \ge 0$, it is easy to see that every finitely supported source $X$ satisfies Cram\'{e}r's condition.
\end{remark}

\subsection{Gaussian Distributions}

Define the Gaussian probability density function and the Gaussian cumulative distribution function as
\begin{align}
\varphi( u )
 \coloneqq
\frac{ 1 }{ \sqrt{ 2 \pi } } \, \mathrm{e}^{-u^{2}/2},\quad\mbox{and}\quad
\Phi( u )
\coloneqq
\int_{-\infty}^{u} \varphi( t ) \, \mathrm{d} t
\end{align}
for $u \in \mathbb{R}$, respectively. Sometimes, we will also find it convenient to use $\mathrm{Q}(u):=1-\Phi(u)$, the Gaussian complementary cumulative distribution function.
Moreover, define
\begin{align}
f_{\mathrm{G}}( s )
& \coloneqq
\begin{cases}
\varphi( \Phi^{-1}( s ) )
& \mathrm{if} \ 0 < s < 1 ,
\\
0
& \mathrm{if} \ s = 0 \ \mathrm{or} \ s = 1 ,
\end{cases}
\label{def:Gaussian_f} \\
g_{\mathrm{G}}( s )
& \coloneqq
\begin{cases}
f_{\mathrm{G}}( \varepsilon ) \, \Phi^{-1}( \varepsilon )
& \mathrm{if} \ 0 < s < 1 ,
\\
0
& \mathrm{if} \ s = 0 \ \mathrm{or} \ s = 1 ,
\end{cases}
\label{def:Gaussian_g}
\end{align}
for $0 \le s \le 1$, where $\Phi^{-1}( \cdot )$ denotes the inverse function of $\Phi( \cdot )$.
It is known that
\begin{align}
\Phi^{-1}( s )
& \sim
- \sqrt{ 2 \ln \frac{ 1 }{ s } } ,
\label{eq:inv-Gauss_1st} \\
f_{\mathrm{G}}( s )
& \sim
s \sqrt{ 2 \ln \frac{ 1 }{ s } }
\label{eq:inv-Gauss_f}
\end{align}
as $s \to 0^{+}$ (cf.\ \cite[Lemma~5.2]{csiszar_korner_2011}), where $\ln$ stands for the natural logarithm.
Thus, we see that
\begin{align}
\lim_{s \to 0^{+}} f_{\mathrm{G}}( \varepsilon ) \, \Phi^{-1}( \varepsilon )
=
- \lim_{s \to 1^{-}} f_{\mathrm{G}}( \varepsilon ) \, \Phi^{-1}( \varepsilon )
=
\lim_{s \to 0^{+}} 2 \, s \ln s
=
0 ,
\end{align}
implying that the definitions of $g_{\mathrm{G}}( s )$ at $s = 0$ and at $s = 1$ are consistent with the limits as $s \to 0^{+}$ and as $s \to 1^{-}$, respectively.
The following lemma shows a higher-order asymptotic expansion of $\Phi^{-1}( \cdot )$ beyond that presented in \eqref{eq:inv-Gauss_1st}.

\begin{lemma}[\cite{blair_edwards_johnson_1976}]
\label{lem:inv-Gauss}
It holds that
\begin{align}
\Phi^{-1}( s )^{2}
=
2 \ln \frac{ 1 }{ 2 \sqrt{\pi} \, s } - \ln \ln \frac{ 1 }{ 2 \sqrt{ \pi } \, s } + \mathrm{O}\left( \frac{ \ln \ln (1/s) }{ \ln (1/s) } \right)
\end{align}
as $s \to 0^{+}$.
\end{lemma}

The following lemma is employed to integrate polynomials of~$\Phi^{-1}( \cdot )$.

\begin{lemma}
\label{lem:int-Gaussian}
Given $0 \le a < b \le 1$, it holds that
\begin{align}
\int_{a}^{b} \Phi^{-1}( s ) \, \mathrm{d} s
& =
f_{\mathrm{G}}( a ) - f_{\mathrm{G}}( b ) ,
\label{eq:int-Gaussian_1} \\
\int_{a}^{b} \Phi^{-1}( s )^{2} \, \mathrm{d} s
& =
(b - a) - g_{\mathrm{G}}(b) + g_{\mathrm{G}}(a) ,
\label{eq:int-Gaussian_2}
\end{align}
\end{lemma}

\begin{IEEEproof}[Proof of \lemref{lem:int-Gaussian}]
Elementary calculations yield these formulas, and we omit the  proof details here.
\end{IEEEproof}

\subsection{Asymptotic Notations}

In this paper, we use the following asymptotic notations to express our asymptotic expansions in source coding problems.
Let $\mathcal{I}_{n}$ be a sequence of real intervals, and $\mathcal{I} = \bigcup_{n} \mathcal{I}_{n}$.
Consider two sequences $\{ f_{n} \}_{n = 1}^{\infty}$ and $\{ g_{n} \}_{n = 1}^{\infty}$ of real-valued functions on $\mathcal{I}$, and a sequence $\{ a_{n} \}_{n = 1}^{\infty}$ of positive numbers.
For fixed $t \in \mathcal{I}$, we say that \emph{$f_{n}(t) = g_{n}(t) + \mathrm{O}( a_{n} )$ as $n \to \infty$} if
\begin{align}
\limsup_{n \to \infty} \frac{ |f_{n}(t) - g_{n}(t)| }{ a_{n} }
<
\infty ,
\end{align}
and that $f_{n}(t) = g_{n}(t) + \mathrm{o}( a_{n} )$ as $n \to \infty$ if
\begin{align}
\lim_{n \to \infty} \frac{ |f_{n}(t) - g_{n}(t)| }{ a_{n} }
=
0 .
\end{align}
In particular, we say that \emph{$f_{n}(t) = g_{n}(t) + \mathrm{O}( a_{n} )$ uniformly on $\mathcal{I}_{n}$ as $n \to \infty$} if
\begin{align}
\limsup_{n \to \infty} \frac{ 1 }{ a_{n} } \sup_{t \in \mathcal{I}_{n}} |f_{n}(t) - g_{n}(t)|
<
\infty ,
\end{align}
and that \emph{$f_{n}(t) = g_{n}(t) + \mathrm{o}( a_{n} )$ uniformly on $\mathcal{I}_{n}$ as $n \to \infty$} if
\begin{align}
\lim_{n \to \infty} \frac{ 1 }{ a_{n} } \sup_{t \in \mathcal{I}_{n}} |f_{n}(t) - g_{n}(t)|
=
0 .
\end{align}
In this study, these uniform convergence properties on a sequence of intervals are used in the moderate deviations analysis to investigate higher-order asymptotics of the fixed-length compression problem in which the error probabilities are asymptotically close to zero or one for sufficiently large codeword lengths; see \sectref{sect:fixed} for details.

\section{Variable-Length Compression}
\label{sect:variable}

\subsection{Variable-Length Compression Allowing Errors---Revisited}

In this subsection, we revisit the previous results stated in \eqref{eq:SV_3rd} and \eqref{eq:KPV_2nd} formally.
Consider compressing a discrete memoryless source $X$ into a finite-length binary string.
Let
\begin{align}
\{ 0, 1 \}^{\ast}
\coloneqq
\{ \varnothing \} \cup \left( \bigcup_{n = 1}^{\infty} \{ 0, 1 \}^{n} \right)
\end{align}
be the set of finite-length binary strings containing the empty string $\varnothing$.
Denote by $\ell : \{ 0, 1 \}^{\ast} \to \mathbb{N} \cup \{ 0 \}$ the length function of a binary string, i.e., $\ell( \varnothing ) = 0$, $\ell( 0 ) = \ell( 1 ) = 1$, $\ell( 00 ) = \ell( 01 ) = \ell( 10 ) = \ell( 11 ) = 2$, etc.

\begin{definition}
An $(L, \varepsilon)$-code for a source $X$ is a pair of a stochastic encoder $F : \mathcal{X} \to \{ 0, 1 \}^{\ast}$ and a stochastic decoder $G : \{ 0, 1 \}^{\ast} \to \mathcal{X}$ such that
\begin{align}
\mathbb{E}[ \ell( F(X) ) ]
& \le
L ,
\\
\mathbb{P}\{ X \neq G(F(X)) \}
& \le
\varepsilon .
\end{align}
\end{definition}

Given a permissible probability of error $0 \le \varepsilon \le 1$, denote by
$L^{\ast}(\varepsilon \mid X)$
the infimum of $L > 0$ such that an $(L, \varepsilon)$-code exists for $X$.
We recall the definition of $\langle \cdot \rangle_{\varepsilon}$ in \eqref{def:cutoff}. It is known (cf.\ \cite[Equation~(26)]{kostina_polyanskiy_verdu_2015} and \cite[Lemmas~1 and~5]{sakai_tan_2019_VL}) that
\begin{align}
L^{\ast}(\varepsilon \mid X^{n})
=
\mathbb{E}[ \langle  \lfloor \log \varsigma_{n}^{-1}( X^{n} ) \rfloor \rangle_{\varepsilon} ] ,
\label{eq:formula_L-ast}
\end{align}
where $\varsigma_{n} : \{ 1, 2, 3, \dots \} \to \mathcal{X}^{n}$ is an arbitrary bijection satisfying%
\footnote{Namely, the bijection $\varsigma_{n}$ plays the role of a decreasing rearrangement of $P_{X^{n}}( \cdot )$.}
\begin{align}
P_{X^{n}}( \varsigma_{n}( 1 ) )
\ge
P_{X^{n}}( \varsigma_{n}( 2 ) )
\ge
P_{X^{n}}( \varsigma_{n}( 3 ) )
\ge
\cdots ,
\end{align}
and $\lfloor \cdot  \rfloor \coloneqq \max\{ z  \in\mathbb{Z}\mid z \le \cdot \}$ denotes the floor function. Since, in the following, we will mostly be working with the inverse $\varsigma_n^{-1}$, for notational conciseness, we write this as  $\gamma_n$, i.e., 
\begin{equation}
\gamma_n(x^n):=\varsigma_n^{-1}(x^n)\quad \mbox{for all}\;\; x^n\in\mathcal{X}^n. 
\end{equation}
Note  that the right-hand side of \eqref{eq:formula_L-ast} is not single-letterized, and we are interested to determine asymptotic expansions of $L^{\ast}(\varepsilon \mid X^{n})$ as $n \to \infty$ in a computable form.

The following theorem is a known second-order asymptotic result for this problem.

\begin{theorem}[{Kostina, Polyanskiy, and Verd\'{u} \cite[Theorem~4]{kostina_polyanskiy_verdu_2015}}]
\label{th:KPV}
Given a fixed $0 \le \varepsilon \le 1$ and a source $X$, it holds that
\begin{align}
L^{\ast}(\varepsilon \mid X^{n})
=
n \, (1 - \varepsilon) \, H(X) - \sqrt{n \, V(X)} \, f_{\mathrm{G}}(\varepsilon) + \mathrm{O}( \log n )
\label{eq:KPV}
\end{align}
as $n \to \infty$, provided that $\mathbb{E}[ \log^{3} P_{X}(X) ]$ is finite.
\end{theorem}

In \cite{kostina_polyanskiy_verdu_2015}, \thref{th:KPV} was proven by establishing the one-shot bounds%
\footnote{When $\varepsilon = 0$, the lower bound specializes Alon and Orlitsky's bound \cite{alon_orlitsky_1994}, and the upper bound specializes Wyner's bound \cite{wyner_1972}.}
\begin{align}
&\mathbb{E}\left[ \left\langle \log \frac{ 1 }{ P_{X^{n}}( X^{n} ) } \right\rangle_{\varepsilon} \right] - \log\Big( 1 + n \, H(X) \Big) - \log \mathrm{e} \nonumber\\*
&\qquad\le
L^{\ast}(\varepsilon \mid X^{n})\le 
\mathbb{E}\left[ \left\langle \log \frac{ 1 }{ P_{X^{n}}( X^{n} ) } \right\rangle_{\varepsilon} \right]
\end{align}
and the asymptotic expansion%
\footnote{This asymptotic expansion was proven by Berry--Esseen-type bounds (cf.\ \cite[Chapter~XVI.5]{feller_1971} and \cite[Chapter~V.4]{petrov_1975}).}
\begin{align}
\mathbb{E}\left[ \left\langle \log \frac{ 1 }{ P_{X^{n}}( X^{n} ) } \right\rangle_{\varepsilon} \right]
=
n (1 \!-\! \varepsilon) H(X) \!-\! \sqrt{ n   V(X) }   f_{\mathrm{G}}( \varepsilon ) \!+\! \mathrm{O}( 1 ) \label{eqn:1minuseps}
\end{align}
as $n \to \infty$.
Roughly speaking, this proof strategy converts the analysis of $\log \gamma_n ( X^{n} )$ to that of the information density $- \log P_{X^{n}}(X^{n})$; see \eqref{eq:formula_L-ast}.
%On the other hand, this proof strategy does not allow us to obtain a  more precise expression of the remainder term $+ \mathrm{O}( \log n )$ of \thref{th:KPV}.

For an asymptotic relation (in an \emph{almost sure} sense) between $\log \gamma_n ( X^{n} )$ and $- \log P_{X^{n}}(X^{n})$ up to the $+\mathrm{o}( \kappa_{n} \log n )$ term with any slowly divergent positive sequence $\{ \kappa_{n} \}_{n = 1}^{\infty}$,  we refer the reader to the study of \emph{pointwise redundancy} studied by Kontoyiannis and Verd\'u~\cite[Section~IV]{kontoyiannis_verdu_2014}.

In the particular case of $\varepsilon = 0$ and finitely supported $X$, \thref{th:KPV} can be refined as follows:

\begin{theorem}[{Szpankowski and Verd\'{u} \cite[Theorem~4]{szpankowski_verdu_2011}}]
\label{th:SV}
For a finitely supported source $X$, it holds that
\begin{align}
L^{\ast}(0 \mid X^{n})
=
n \,  H(X) - \frac{ 1 }{ 2 } \log n + \mathrm{O}( 1 )
\label{eq:SV}
\end{align}
as $n \to \infty$.
\end{theorem}

In \cite{szpankowski_verdu_2011}, \thref{th:SV} was proven via complex analysis so-called the \emph{analytic Poissonization and de-Poissonization} (cf.\ \cite{szpankowski_average}) and Stirling's formula to approximate multinomial coefficients.

%\begin{remark}
%Recently, the present authors \cite{sakai_tan_2019_VL} investigated the problem considering variable-length compression allowing errors in the case when some side-information $Y$ of the source $X$ is available at both encoder and decoder.
%For this problem, we introduced two error formalisms: the average and maximum error criteria, where the averaging and maximization are taken with respect to $Y$.
%We showed that the first-order optimal coding rates are the same under both error criteria, and the second-order optimal coding rates differ under these error criteria.
%In particular, this difference can be characterized by the law of total variance for the conditional information density $- \log P_{X|Y}(X \mid Y)$.
%\end{remark}

\subsection{Main Result---Higher-Order Asymptotics of Variable-Length Compression}
\label{sect:main}

The following theorem 
%is a refinement of Theorems~\ref{th:KPV} and~\ref{th:SV} and 
constitutes the main result of the paper. 

\begin{theorem}
\label{th:third}
Let $0 < \varepsilon \le 1$ be fixed.
If the source $X$ satisfies Cram\'{e}r's condition, then
\begin{align}
&L^{\ast}(\varepsilon \mid X^{n})\nonumber\\*
&=
n \, (1 -\varepsilon)  \,  H(X)- \sqrt{ n \,  V(X) }   f_{\mathrm{G}}( \varepsilon )-\frac{ 1 -\varepsilon }{ 2 } \log n + \mathrm{O}( 1 )
\label{eq:third}
\end{align}
as $n \to \infty$.
On the other hand, if $\varepsilon = 0$ and $X$ is finitely supported, then \eqref{eq:SV} holds.
\end{theorem}
 Our main result in~\eqref{eq:third} provides a third-order refinement of Kostina, Polyanskiy, and Verd\'{u}'s second-order result in~\eqref{eq:KPV_2nd} for variable-length lossless compression with errors \cite[Theorem~4]{kostina_polyanskiy_verdu_2015}.

We prove \thref{th:third} in the next section. %, which contains also an alternative proof of \thref{th:SV}.
%Our proof strategy relies more heavily on information-spectrum methods \cite{han_2003} compared to the original proof in \cite[Section~V]{szpankowski_verdu_2011}.

%\begin{remark}
%Note that the third-order term in \eqref{eq:third}, being $- ((1 - \varepsilon)/2) \log n$, differs from the usual third-order terms for other information-theoretic problems such as channel  and source coding (cf.\ \cite{tan_2014}), in which these terms do not depend on $\varepsilon$.
%\end{remark}

The proof outline of \thref{th:third} is as follows:
Since every codeword length is a nonnegative integer, it is known that
\begin{align}
L^{\ast}(0 \mid X^{n})
=
\sum_{k = 1}^{\infty} \mathbb{P}\{ \log \gamma_n ( X^{n} ) \ge k \}
\end{align}
(cf.\ \cite[Section~III]{kontoyiannis_verdu_2014}).
Given $0 < \varepsilon \le 1$, this identity can be readily extended as
\begin{align}
L^{\ast}(\varepsilon \mid X^{n})
=
\sum_{k = 1}^{\tilde{\xi}_{n}} \mathbb{P}\{ \log \gamma_n ( X^{n} ) \ge k \} - \varepsilon \, \tilde{\xi}_{n} ,
\end{align}
where the integer $\tilde{\xi}_{n} = \tilde{\xi}_{n}(\varepsilon, X)$ is chosen so that
\begin{align}
\mathbb{P}\{ \log \gamma_n ( X^{n} ) \ge \tilde{\xi}_{n} \}
& \ge
\varepsilon ,
\label{eq:xi_tilde_1} \\*
\mathbb{P}\{ \log \gamma_n ( X^{n} ) > \tilde{\xi}_{n} \}
& <
\varepsilon .
\label{eq:xi_tilde_2}
\end{align}
Here, the complementary cumulative distribution function $\mathbb{P}\{ \log \gamma_n ( X^{n} ) > k \}$ corresponds to the \emph{overflow probability of codeword lengths} (cf.\ \cite{merhav_1991, kontoyiannis_verdu_2014, kosut_sankar_2017, saito_matsushima_2017, iri_kosut_2019, nomura_yagi_2019}), and can be thought of as the minimum average probability of error for \emph{$n$-to-$k$ binary block codes for the source $X^{n}$.}
Namely, the average codeword length $L^{\ast}(\varepsilon \mid X^{n})$ of \emph{variable-length} compression can be analyzed via the fundamental limits of \emph{fixed-length} compression via its relation to  $\log\gamma_n(X^n)$ as stated in~\eqref{eq:formula_L-ast}.%
\footnote{This relation was mentioned by S.\ Verd\'{u} in his Shannon Lecture \cite{verdu_isit2007}.} In particular in Lemma \ref{lem:cutoff-log_integral-M}, we show that  the expectation  of the $\varepsilon$-cutoff transformation action on $\log\gamma_n(X^n)$ (with the floor operator removed) %, namely $\langle  \lfloor \log\gamma_n( X^{n} ) \rfloor \rangle_{\varepsilon}$ in~\eqref{eq:formula_L-ast} 
is related to the fixed-length source coding asymptotics $ M^*(n,s)$ as follows:
\begin{equation}
\mathbb{E}\left[ \langle\log\gamma_n(X^n)  \rangle_\varepsilon \right]\approx\int_\varepsilon^1\log M^*(n,s)\,\mathrm{d}s. 
\end{equation}
To bound the integral on the right-hand-side, we split it into two parts (as error probabilities close to $0$ or $1$ are difficult to deal with using central limit-type techniques), namely
\begin{align}
&\int_\varepsilon^1\log M^*(n,s)\,\mathrm{d}s\nonumber\\*
&=\underbrace{\int_{\varepsilon}^{1-n^{-1}}\log M^*(n,s)\,\mathrm{d}s}_{=:\mathrm{A}} + \underbrace{\int_{1-n^{-1}}^{1}\log M^*(n,s)\,\mathrm{d}s }_{=: \mathrm{B}}.
\end{align}
Integral $\mathrm{B}$ only contributes an $\mathrm{O}(1)$ term. Estimating integral $\mathrm{A}$ is the essence of the whole proof. It requires us to estimate $\log M^*(n,\varepsilon_n)$ for error probabilities $\varepsilon_n$ that are vanishing or growing polynomially fast, e.g., $\varepsilon_n\approx n^{-1}$ or $\varepsilon_n \approx 1-n^{-1}$. Thus, one of our main endeavors  and contributions (done in~\lemref{lem:MD}) is to estimate $\log M^*(n,1/n^r)$ and $\log M^*(n,1-1/n^r)$ for $r>0$, i.e., the fundamental limits of fixed-length source coding when the error probability or success probability is polynomially small. This requires techniques from the theory of moderate deviations and strong large deviations.

\section{Proof  of \thref{th:third}}
\label{sect:fourth}

In this section, we prove \thref{th:third} by presenting some technical lemmas.

\subsection{Moderate Deviations and Strong Large Deviations of Fixed-Length Compression}
\label{sect:fixed}

Before investigating higher-order asymptotic expansions of the variable-length compression problem, we now consider the \emph{fixed-length} compression problem for a stationary memoryless source $X^{n}$.
An \emph{$(n, M, \varepsilon)$-code for the source $X$} consists of  an encoder $f : \mathcal{X}^{n} \to \{ 1, 2, \dots, M \}$ and a decoder $g : \{ 1, 2, \dots, M \} \to \mathcal{X}^{n}$ such that
\begin{align}
\mathbb{P}\{ X^{n} \neq g(f(X^{n})) \}
\le
\varepsilon .
\end{align}
Denote by $M^{\ast}(n, \varepsilon)$ the minimum of $M \in \mathbb{N}$ such that an $(n, M, \varepsilon)$-code exists for the source $X$.
In other words, it is defined as
\begin{align}
M^{\ast}(n, \varepsilon)
=
\min_{\substack{ \mathcal{A} \subset \mathcal{X} : \\ P_{X^n}(\mathcal{A}) \ge 1 - \varepsilon }} |\mathcal{A}| .
\label{eq:HT}
\end{align}

The following lemma is a result of judiciously combining the use of a  \emph{moderate deviations theorem} and a \emph{strong large deviations theorem} \cite{feller_1971, petrov_1975, dembo_zeitouni_1998}. %\red{In the following, let $\mu_3= \mathbb{E}[ ( -\ln P_X(X)-H(X))^3]$ be the third central moment of the entropy density $-\ln P_X(X)$  and $\cX = \mu_3/(6V(X)^{3/2})$. }

\begin{lemma}
\label{lem:MD_block}
Suppose that $X$ satisfies Cram\'{e}r's condition stated in \defref{def:Cramer}.
Let $\{ \varepsilon_{n} \}_{n = 1}^{\infty}$ be a real sequence on $(0, 1)$.
If
\begin{align}
\frac{ 1 }{ n^{r} }
\le
\varepsilon_{n}
\le
1 - \frac{ 1 }{ n^{r} }
\label{eq:polynomial_eps}
\end{align}
for some positive real $r$ and for sufficiently large $n$. Then%
\footnote{The remainder term $+ \mathrm{O}( 1 )$ in \eqref{eq:MD_block} depends only on $X$ and $r$, i.e., it is independent of the sequence $\{ \varepsilon_{n} \}_{n = 1}^{\infty}$.}
\begin{align}
\log M^{\ast}(n, \varepsilon_{n})
&=
n \, H(X) - \sqrt{ n \, V(X) } \, \Phi^{-1}( \varepsilon_{n} )  - \frac{ 1 }{ 2 } \log n \nonumber\\*
&\qquad+ \left(  \cX -\frac{\log \mathrm{e}}{ 2 }\right) \Phi^{-1}( \varepsilon_{n} )^{2} + \mathrm{O}( 1 )
\label{eq:MD_block}
\end{align}
as $n \to \infty$, where $\cX$ is defined in \eqref{eqn:skewness}.
\end{lemma}

\begin{IEEEproof}[Proof of \lemref{lem:MD_block}]
See \sectref{sec:proofLem3}.
\end{IEEEproof}

\begin{remark}[Refinements to the source dispersion term in the moderate deviations regime]
Define
\begin{align}
D^{\ast}(n, \varepsilon)
\coloneqq
\frac{ \log M^{\ast}(n, \varepsilon) - n \, H(X) }{ \sqrt{ n \, V(X) } } .
\end{align}
It is well-known that $D^{\ast}(n, \varepsilon) \to \Phi^{-1}( 1 - \varepsilon )$ as $n \to \infty$ for fixed $0 < \varepsilon < 1$.
More precisely, it is clear from \eqref{eq:third_fixed} that
\begin{align}
D^{\ast}(n, \varepsilon)
=
\Phi^{-1}(1 - \varepsilon) - \frac{ \log n }{ 2 \sqrt{ n \, V(X) } } + \mathrm{O}\left( \frac{ 1 }{ \sqrt{n} } \right) 
\label{eq:source-dispersion}
\end{align}
as $n \to \infty$ for fixed $0 < \varepsilon < 1$.
By \lemref{lem:MD_block}, Equation~\eqref{eq:source-dispersion} can be extended to the case when $\varepsilon$ approaches to either zero or one polynomially in $n$ as follows:
Given an arbitrary positive real number $r$, it follows from Lemmas~\ref{lem:inv-Gauss} and~\ref{lem:MD_block} that
\begin{align}
D^{\ast}(n, n^{-r})
& =
\Phi^{-1}( 1 - n^{-r} ) - \frac{ (1 + 2 \, r) \log n - \log \log n }{ 2 \sqrt{ n \, V(X) } }  \nonumber\\*  
&\qquad+\mathrm{O}\left( \frac{ 1 }{ \sqrt{n} } \right) ,
\label{eq:source-dispersion_MD1} \\
D^{\ast}(n, 1 - n^{-r})
& =
\Phi^{-1}( n^{-r} ) - \frac{ (1 + 2 \, r) \log n - \log \log n }{ 2 \sqrt{ n \, V(X) } } \nonumber\\*  
&\qquad+ \mathrm{O}\left( \frac{ 1 }{ \sqrt{n} } \right)
\label{eq:source-dispersion_MD2}
\end{align}
as $n \to \infty$.
To asymptotically expand $\Phi^{-1}(1 - n^{-r})$ and $\Phi^{-1}( n^{-r} )$ in \eqref{eq:source-dispersion_MD1} and \eqref{eq:source-dispersion_MD2}, respectively, we see from \lemref{lem:inv-Gauss} that
\begin{align}
D^{\ast}(n, n^{-r})^{2}
& =
\Phi^{-1}( 1 - n^{-r} )^{2} + \mathrm{O}\left( \frac{ \log^{3/2} n }{ \sqrt{n} } \right) \nonumber\\*
&=
2 \, r \ln n - \ln \frac{ \pi }{ 2 } - \ln \Big( 2 \, r \ln n - \ln \pi \Big)  \nonumber\\* 
&\qquad+ \mathrm{O}\left( \frac{ \ln \ln n }{ \ln n } \right)
\label{eq:source-dispersion_MD3}
\end{align}
as $n \to \infty$; and analogously, we get
\begin{align}
D^{\ast}(n, 1 - n^{-r})^{2}
& =
\Phi^{-1}( n^{-r} )^{2} + \mathrm{O}\left( \frac{ \log^{3/2} n }{ \sqrt{n} } \right)\nonumber\\*
&=
2 \, r \ln n - \ln \frac{ \pi }{ 2 } - \ln \Big( 2 \, r \ln n - \ln \pi \Big) \nonumber\\* 
&\qquad+ \mathrm{O}\left( \frac{ \ln \ln n }{ \ln n } \right)
\label{eq:source-dispersion_MD4}
\end{align}
as $n \to \infty$.
Thus, we have obtained expressions for the  higher-order optimal coding rates of the fixed-length compression problem when the error probabilities vanish polynomially in the blocklength $n$. 
%In the classical-quantum channel coding problem, the evaluation of the pre-factors in moderate deviations theorem was posed in \cite[Section~V]{chubb_tan_tomamichel_2017} as an avenue for  future work.
%Note that the remainder terms $+ \mathrm{O}( (\ln \ln n) / (\ln n) )$ in the right-hand sides of \eqref{eq:source-dispersion_MD3} and \eqref{eq:source-dispersion_MD4} can be further refined by using higher-order asymptotic expansions of $u \mapsto \Phi( u )$ as $u \to -\infty$ beyond that stated in~\lemref{lem:inv-Gauss}; see \cite{blair_edwards_johnson_1976} and \cite[Chapter~VII.7]{feller_1968}.
\end{remark}

Our techniques to prove Lemma \ref{lem:MD_block} involve using newly-developed moderate deviations results in \sectref{sect:MD} and strong large deviations results in \sectref{sect:SLD} to analyze the fixed-length compression problem.
These result  in a new asymptotic expansion for fixed-length compression in \lemref{lem:MD_block} that is also amenable to integration over the error probability parameter (over a certain range) to obtain a third-order asymptotic expansion for the variable-length compression problem.

\subsection{On the Cutoff Operation for Logarithm of Integer-Valued Random Variable}
\label{sect:cutoff-log}

%Due to the floor function in the expectation on the right-hand side of \eqref{eq:formula_L-ast}, it is difficult to deal with the exact fourth-order term of $L^{\ast}(\varepsilon \mid X^{n})$, i.e., the $+\mathrm{O}( 1 )$ term of $L^{\ast}(\varepsilon \mid X^{n})$.
%Instead, 

We now investigate a simplified version of the expectation in~\eqref{eq:formula_L-ast}. The following two lemmas derive asymptotic expressions for $\mathbb{E}[ \langle \log \gamma_n ( X^{n} ) \rangle_{\varepsilon} ]$, i.e., the expectation of $\langle \log \gamma_n ( X^{n} ) \rangle_{\varepsilon}$  in the absence of the floor function  noting that this  operation  does  not affect the asymptotics.

\begin{lemma}
\label{lem:cutoff-log_integral-M}
Given $0 \le \varepsilon \le 1$, it holds that
\begin{align}
\mathbb{E}[ \langle \log \gamma_n ( X^{n} ) \rangle_{\varepsilon} ]
=
\int_{\varepsilon}^{1} \log M^{\ast}(n, s) \, \mathrm{d} s + \mathrm{o}( 1 )
\label{eq:cutoff-log_integral-M}
\end{align}
as $n \to \infty$.
\end{lemma}

\begin{IEEEproof}[Proof of \lemref{lem:cutoff-log_integral-M}]
See \appref{app:cutoff-log_integral-M}.
\end{IEEEproof}

\begin{remark}
The quantity $\mathbb{E}[ \langle \log \gamma_n ( X^{n} ) \rangle_{\varepsilon} ]$ that appears on  the left-hand side of \eqref{eq:cutoff-log_integral-M} is closely related to the fundamental limits of the guessing problem \cite{massey_isit1994, arikan_1996} allowing errors \cite{kuzuoka_2020} for a source $X$; see \cite[Section~IV]{sakai_tan_2019_VL}.
\end{remark}

\begin{remark}\label{rem:SV}
In \cite[Equations~(44)--(48)]{szpankowski_verdu_2011}, Szpankowski and Verd\'{u} showed that
\begin{align}
\mathbb{E}[ \lfloor \log \gamma_n ( X^{n} ) \rfloor ]
=
\int_{0}^{1} \lceil \log M^{\ast}(n, s) \rceil \, \mathrm{d} s - 1 ,
\label{eq:SV_identity}
\end{align}
and the proof of \lemref{lem:cutoff-log_integral-M} is similar to that of this identity.
Note that $\lceil \log M^{\ast}(n, \varepsilon) \rceil$ denotes the infimum of integers $k$ such that an $n$-to-$k$ binary block code for which the error probability is at most $\varepsilon$  exists.
This quantity is slightly different from the fixed-length compression problem described in \sectref{sect:fixed}.
\end{remark}

\begin{lemma}
\label{lem:exp-log}
Let $0 < \varepsilon \le 1$ be fixed.
If the source $X$ satisfies Cram\'{e}r's condition, then
\begin{align}
&\mathbb{E}[ \langle \log \gamma_n ( X^{n} ) \rangle_{\varepsilon} ]\nonumber\\* 
&
=
n \, (1 - \varepsilon) \, H(X) - \sqrt{ n \, V(X) } \, f_{\mathrm{G}}(\varepsilon) - \frac{ 1 - \varepsilon }{ 2 } \log n + \mathrm{O}( 1 )
\label{eq:exp-log_MD}
\end{align}
as $n \to \infty$.
On the other hand, if $X$ is finitely supported, then
\begin{align}
\mathbb{E}[ \log \gamma_n ( X^{n} ) ]
& =
n \, H(X) - \frac{ 1 }{ 2 } \log n + \mathrm{O}( 1 )
\label{eq:exp-log_finite}
\end{align}
as $n \to \infty$.
\end{lemma}

\begin{IEEEproof}[Proof of \lemref{lem:exp-log}]
The idea of the proof of \lemref{lem:exp-log} is to evaluate the integral in  \lemref{lem:cutoff-log_integral-M}. For this purpose,  we  suppose that $X$ satisfies Cram\'{e}r's condition.
Since $\varepsilon \mapsto \log M^{\ast}(n, \varepsilon)$ is nonnegative and nonincreasing on $(0, 1)$, we readily see that
\begin{align}
&\int_{\varepsilon}^{1 - n^{-1}} \log M^{\ast}(n, s) \, \mathrm{d} s
\le
\int_{\varepsilon}^{1} \log M^{\ast}(n, s) \, \mathrm{d} s\nonumber\\* 
&\qquad
\le
\int_{\varepsilon}^{1 - n^{-1}} \log M^{\ast}(n, s) \, \mathrm{d} s + \frac{ 1 }{ n } \log M^{\ast}(n, \varepsilon)
\label{eq:cut_tail_int_1}
\end{align}
for $n \ge (1 - \varepsilon)^{-1}$. Define 
\begin{align}
&K_{n}(X) \nonumber\\* 
&\coloneqq
\max_{n^{-1} \le \varepsilon \le 1 - n^{-1}} \bigg| \log M^{\ast}(n, \varepsilon) - \bigg( n \,   H(X) - \sqrt{ n \, V(X) } \, \Phi^{-1}( \varepsilon ) \nonumber\\* 
&\qquad\qquad
- \frac{ 1 }{ 2 } \log n +\Big(\cX-\frac{\log \mathrm{e}}{2}\Big) \, \Phi^{-1}( \varepsilon )^{2} \bigg) \bigg| .
\end{align}
Now, \lemref{lem:MD_block} implies that
\begin{align}
K_{n}(X)
=
\mathrm{O}( 1 )
\label{eq:K_MD_block}
\end{align}
as $n \to \infty$. 
%, where
%\begin{align}
%K_{n}(X)
%\coloneqq
%\max_{n^{-1} \le \varepsilon \le 1 - n^{-1}} \left| \log M^{\ast}(n, \varepsilon) - \left( n \, (1 - \varepsilon) \, H(X) - \sqrt{ n \, V(X) } \, \Phi^{-1}( \varepsilon ) - \frac{ 1 }{ 2 } \log n - \frac{ \Phi^{-1}( \varepsilon )^{2} }{ 2 } \right) \right| .
%\end{align}
Then, it follows from \lemref{lem:int-Gaussian} and~\eqref{eq:K_MD_block} that
\begin{align}
&\int_{\varepsilon}^{1 - n^{-1}} \log M^{\ast}(n, s) \, \mathrm{d} s\nonumber\\* 
& \le
n \left( 1 - \frac{ 1 }{ n } - \varepsilon \right) H(X) - \sqrt{ n \, V(X) } \, \Big( f_{\mathrm{G}}( \varepsilon ) - f_{\mathrm{G}}( n^{-1} ) \Big)
\notag \\
& \qquad {}
- \frac{ 1 - n^{-1} - \varepsilon }{ 2 } \log n \nonumber\\* 
&\qquad
- \frac{ (\log \mathrm{e})(1 - n^{-1} - \varepsilon - g_{\mathrm{G}}( 1 - n^{-1} ) + g_{\mathrm{G}}( \varepsilon )) }{ 2  } + K_{n}(X)
\notag \\
& =
n  (1 - \varepsilon)  H(X) - \sqrt{ n   V(X) }   f_{\mathrm{G}}( \varepsilon ) - \frac{ 1 - \varepsilon }{ 2 } \log n + \mathrm{O}( 1 )
\label{eq:int_cutting-tail_1a}
\end{align}
as $n \to \infty$.
Analogously, we get
\begin{align}
&\int_{\varepsilon}^{1 - n^{-1}} \log M^{\ast}(n, s) \, \mathrm{d} s\nonumber\\* 
& \ge
n \left( 1 - \frac{ 1 }{ n } - \varepsilon \right) H(X) - \sqrt{ n \, V(X) } \, \Big( f_{\mathrm{G}}( \varepsilon ) - f_{\mathrm{G}}( n^{-1} ) \Big)
\notag \\
& \qquad {}
- \frac{ 1 - n^{-1} - \varepsilon }{ 2 } \log n \nonumber\\* 
&\qquad
- \frac{ (\log \mathrm{e}) (1 - n^{-1} - \varepsilon - g_{\mathrm{G}}( 1 - n^{-1} )  + g_{\mathrm{G}}( \varepsilon) ) }{ 2  } - K_{n}(X)
\notag \\
& =
n (1 - \varepsilon)   H(X) - \sqrt{ n  V(X) }  f_{\mathrm{G}}( \varepsilon ) - \frac{ 1 - \varepsilon }{ 2 } \log n + \mathrm{O}( 1 )
\label{eq:int_cutting-tail_1b}
\end{align}
as $n \to \infty$.
Since $\log M^{\ast}(n, \varepsilon) = \mathrm{O}( n )$ as $n \to \infty$, combining \eqref{eq:cut_tail_int_1}, \eqref{eq:int_cutting-tail_1a}, and \eqref{eq:int_cutting-tail_1b}, we obtain \eqref{eq:exp-log_MD} of \lemref{lem:exp-log}.

Finally, suppose that $X$ is finitely supported.
Similar to \eqref{eq:cut_tail_int_1}, we get
\begin{align}
&\int_{n^{-1}}^{1 - n^{-1}} \log M^{\ast}(n, s) \, \mathrm{d} s
 \le
\int_{0}^{1} \log M^{\ast}(n, s) \, \mathrm{d} s
\notag \\*
& \le
\int_{n^{-1}}^{1 - n^{-1}} \log M^{\ast}(n, s) \, \mathrm{d} s + 2 \log |\operatorname{supp}(X)| .
\label{eq:cut_tail_int_2}
\end{align}
Since every finitely supported source $X$ satisfies Cram\'{e}r's condition, it follows from Lemmas~\ref{lem:int-Gaussian} and~\ref{lem:MD_block} that
\begin{align}
&\int_{n^{-1}}^{1 - n^{-1}} \log M^{\ast}(n, s) \, \mathrm{d} s  =
n \left( 1 - \frac{ 2 }{ n } \right) H(X) - \frac{ 1 - 2 \, n^{-1} }{ 2 } \log n\nonumber\\* 
&\qquad - \frac{ (\log \mathrm{e})(1 - 2 \, n^{-1} + 2 \, g_{\mathrm{G}}( n^{-1} )) }{ 2} + \mathrm{O}( 1 )
\notag \\
& =
n \, H(X) - \frac{ 1 }{ 2 } \log n + \mathrm{O}( 1 )
\label{eq:int_cutting-tail_2}
\end{align}
as $n \to \infty$.
Combining \eqref{eq:cut_tail_int_2} and \eqref{eq:int_cutting-tail_2}, we obtain \eqref{eq:exp-log_finite} of \lemref{lem:exp-log}.
This completes the proof of \lemref{lem:exp-log}.
\end{IEEEproof}

We now use the above to complete the proof of~\thref{th:third}. We    see from \eqref{eq:formula_L-ast} that (taking the floor operation into account)
\begin{align}
\big| \mathbb{E}[ \langle \log \gamma_n ( X^{n} ) \rangle_{\varepsilon} ] - L^{\ast}(\varepsilon \mid X^{n})\big|\le 1.
%=
%\mathbb{E}[ \langle \{ \log \gamma_n ( X^{n} ) \} \rangle_{\varepsilon} ] .
\label{eq:formula_fractional-part}
\end{align}
%where $\{ u \} \coloneq u - \lfloor u \rfloor$ denotes the fractional part of $u \in \mathbb{R}$.
%Thus, since $0 \le \{ u \} < 1$ for every $u \in \mathbb{R}$, it is clear that the gap between $L^{\ast}(\varepsilon \mid X^{n})$ and $\mathbb{E}[ \langle \log \gamma_n ( X^{n} ) \rangle_{\varepsilon} ]$ is at most $1$ bit.
Therefore, \lemref{lem:exp-log} implies \thref{th:third}, completing the proof.
\hfill\IEEEQEDhere

\section{Higher-Order Asymptotics of Fixed-Length Compression}
\label{sect:block}

In this section, we prove \lemref{lem:MD_block} by employing certain variants of the moderate deviations and strong large deviations theorems.
In the next two subsections, we introduce these fundamental results.

\subsection{Moderate Deviations}
\label{sect:MD}

Consider i.i.d.\ copies $\{ Z_{i} \}_{i = 1}^{\infty}$ of a real-valued r.v.\ $Z$ with zero mean.
Suppose that the variance of $Z$,
\begin{align}
\sigma^{2}
& \coloneqq
\mathbb{E}[ Z^{2} ] ,
\end{align}
is positive and finite.
Now, we want to characterize the distribution function defined by
\begin{align}
F_{n}( z )
\coloneqq
\mathbb{P}\left\{ \sum_{i = 1}^{n} Z_{i} \le z \, \sigma \sqrt{n} \right\}
\end{align}
for each $z \in \mathbb{R}$.
The \emph{central limit theorem} states that
\begin{align}
F_{n}( z )
=
\Phi( z ) + \mathrm{o}( 1 )
\label{eq:CLT}
\end{align}
uniformly on $\mathbb{R}$ as $n \to \infty$.
In this study, to examine higher-order asymptotics of source coding problems either with \emph{vanishing error probabilities} or with \emph{vanishing correct probabilities,} we shall control the error term in \eqref{eq:CLT} more precisely when $z$ diverges as $n \to \infty$.
To this end, we shall use the following version of the \emph{moderate deviations theorem.}

\begin{lemma}[{\cite[Chapter~VIII.2]{petrov_1975}}]
\label{lem:refined_MD}
%\red{Let $\cZ := \mu_3/(6\sigma^3)$ where $\mu_3=\mathbb{E}[ (-\ln P_Z(Z)-H(Z))^3 ]$ is the third central moment of the entropy density $-\ln P_Z(Z)$.} 
Suppose that the moment-generating function $\mathbb{E}[ \mathrm{e}^{t Z} ]$ is finite for some neighborhood of $t = 0$ (i.e., Cram\'{e}r's condition on $Z$).
Given a nonnegative real sequence $\{ z_{n} \}_{n = 1}^{\infty}$ satisfying $z_{n} = \mathrm{O}( n^{1/6} )$ as $n \to \infty$, it holds that
\begin{align}
1 - F_{n}( z_{n} )
& =
\Big( 1 \!-\! \Phi( z_{n} ) \Big) \, \exp\left( \frac{\cZ \, z_{n}^{3} }{  \sqrt{n} } \right) + \mathrm{O}\left( \frac{ \varphi( z_{n} ) }{ \sqrt{n} } \right) ,\label{eqn:FnZn0}
\\
F_{n}( - z_{n} )
& =
\Phi( - z_{n} ) \, \exp\left( - \frac{\cZ \, z_{n}^{3} }{ \sqrt{n}} \right) + \mathrm{O}\left( \frac{ \varphi( z_{n} ) }{ \sqrt{n} } \right) \label{eqn:FnZn}
\end{align}
as $n \to \infty$, where $\cX$ (one-sixth of the skewness) was defined in \eqref{eqn:skewness}. 
\end{lemma}

Given a real number $0 < \varepsilon < 1$, choose $\zeta_{n}(\varepsilon) \in \mathbb{R}$ so that
\begin{align}
\zeta_{n}(\varepsilon)
\coloneqq
\inf\{ z \in \mathbb{R} \mid F_{n}( z ) \ge 1 - \varepsilon \} .
\label{def:zeta}
\end{align}
By \eqref{eq:CLT}, one readily sees
\begin{align}
\Phi( \zeta_{n}( \varepsilon ) )
=
1 - \varepsilon + \mathrm{o}( 1 )
\label{eq:ordinary-CLT}
\end{align}
uniformly on $(0, 1)$ as $n \to \infty$.
We will, however, require a statement similar to \eqref{eq:ordinary-CLT} when $\varepsilon$ is a \emph{sequence} $\{ \varepsilon_{n} \}_{n = 1}^{\infty}$ with limit infimum and limit supremum respectively equal to zero and one.
The sequence $\{ \varepsilon_n \}_{n=1}^\infty$ should also have the property that its  subsequences  approach zero or one polynomially fast.
In fact, we will require a stronger statement that also quantifies the ``rate of convergence''. 

Essentially, we are interested in ``inverting'' the moderate deviations result in Lemma~\ref{lem:refined_MD}. That is, suppose that $F_n(-z_n) = \varepsilon_n$ where $F_n(-z_n)$ is the quantity having asymptotic expansion in \eqref{eqn:FnZn}, we would like to find how $z_n$ scales as $n\to\infty$. By the central limit theorem, we know that $z_n = -\Phi^{-1} ( \varepsilon_n(1+\mathrm{o}(1))$. We would like to carefully obtain a refinement of the $\mathrm{o}(1)$ term. In our application of these results, we will take $\varepsilon_n$ to be either $1/n$ or $1-1/n$ and thus, $z_n\approx \pm\sqrt{2\ln n}$. 
Using \lemref{lem:refined_MD}, we refine the $+\mathrm{o}(1)$ term in \eqref{eq:ordinary-CLT} multiplicatively  as follows:

\begin{lemma}
\label{lem:MD}
 Let $\{ \varepsilon_{n} \}_{n = 1}^{\infty}$ be a real sequence satisfying\footnote{In our application of this result, we will take $\varepsilon_n=1/n$ which satisfies \eqref{eq:eps_small_range} with $r=1$.}
\begin{equation}
\frac{1}{n^r}\le\varepsilon_{n}\le\frac{1}{2}\label{eq:eps_small_range}. 
\end{equation}
for some positive constant $r$. %Recall the definition of $d$ from Lemma~\ref{lem:refined_MD}.
Suppose that the moment generating function $\mathbb{E}[ \mathrm{e}^{t Z} ]$ is finite for some neighborhood of $t = 0$.
Then, it holds that
\begin{align}
1 - \Phi( \zeta_{n}(\varepsilon_{n}) ) &\!=\! \varepsilon_n \left( 1 \!+\! \frac{\cZ  \Phi^{-1} (\varepsilon_n)^3 }{\sqrt{n}}\!-\!\mathrm{O}\bigg( \frac{\Phi^{-1} (\varepsilon_n)}{\sqrt{n}} \bigg) \right) \label{eq:MD_1} \\
 \Phi( \zeta_{n}(1-\varepsilon_{n}) ) &\!=\! \varepsilon_n \left( 1 \! -\!\frac{\cZ \Phi^{-1} (\varepsilon_n)^3}{\sqrt{n}}\!+\!\mathrm{O}\bigg( \frac{\Phi^{-1} (\varepsilon_n)}{\sqrt{n}} \bigg) \right) \label{eq:MD_2}
\end{align}
as $n \to \infty$, where $\cZ$ (one-sixth of the skewness) was defined in \eqref{eqn:skewness}. 
\end{lemma}

\begin{IEEEproof}[Proof of \lemref{lem:MD}]
See \appref{app:MD}.
\end{IEEEproof}

\begin{remark}
Given a finitely supported nonlattice source $X$, let $Z = - \ln P_{X}( X )$.
Then, for fixed $0 < \varepsilon < 1$, asymptotic expansions
\begin{align}
\Phi( \zeta_{n}( \varepsilon ) )
& =
1 - \varepsilon + \mathrm{O}( n^{-1/2} ) ,
\label{eq:CLT_strassen}
\end{align}
as $n \to \infty$ were investigated by Strassen \cite[Equation~(2.21)]{strassen_1964} based on the Edgeworth expansion (cf.\ \cite[Chapter~XVI.4]{feller_1971} or \cite[Chapter~VI.3]{petrov_1975}).
For a precise analysis of the Berry--Esseen bound used to derive \eqref{eq:CLT_strassen}, we refer the reader to Kontoyiannis and Verd\'{u}'s work \cite[Section~V]{kontoyiannis_verdu_2014}.
On the other hand, \lemref{lem:MD} exhibits similar asymptotic expansions when either $\varepsilon_{n} \to 0^{+}$ or $\varepsilon_{n} \to 1^{-}$ along certain subsequences polynomially as $n \to \infty$.
\end{remark}

\subsection{Strong Large Deviations}
\label{sect:SLD}

In this subsection, we introduce strong large deviations theorems for \emph{$\sigma$-finite measures} that are not necessarily probability measures.%
\footnote{For strong large deviations for \emph{finite measures,} refer to \cite[Section~VIII]{hayashi_2018} or \cite[Footnote~8]{tan_hayashi_2018}.
In this study, we consider $\sigma$-finite measures to deal with a countably infinite source alphabet $\mathcal{X}$, because the results on finite measures are applicable only for finite source alphabets.}
Let $(\Omega, \mathcal{F}, \mu)$ be a $\sigma$-finite measure space, and $f : \Omega \to \mathbb{R}$ a Borel-measurable function.
Denote by $\mu_{f} \coloneqq \mu \circ f^{-1}$ the measure on $\mathbb{R}$ induced by $f$.
Define the cumulant generating function as
\begin{align}
\Lambda_{\mu_{f}}(s)
\coloneqq
\ln \left( \int_{\mathbb{R}} \mathrm{e}^{s \, t} \, \mu_{f}( \mathrm{d} t )  \right) ,
\end{align}
and the Fenchel--Legendre transform of $\Lambda_{\mu_{f}}( s )$ by
\begin{align}
\Lambda_{\mu_{f}}^{\ast}( a )
\coloneqq
\sup_{s \in \mathbb{R}} \Big( a \, s - \Lambda_{\mu_{f}}( s ) \Big) .
\end{align}

Let $\mathcal{D}_{\mu_{f}} \coloneqq \{ s \mid \Lambda_{\mu_{f}}( s ) < \infty \}$ and $\operatorname{\mathrm{int}}( \mathcal{D}_{\mu_{f}} )$ its interior.
Similar to \cite[Lemma~2.2.5 and Exercise~2.2.24]{dembo_zeitouni_1998}, it can be verified by H\"{o}lder's inequality and the dominated convergence theorem for the Lebesgue integrals with respect to a $\sigma$-finite measure that $\Lambda_{\mu_{f}}(s)$ is of class $C^{\infty}$ in $s \in \operatorname{\mathrm{int}}( \mathcal{D}_{\mu_{f}} )$.
Especially, it holds that for each $s \in \operatorname{\mathrm{int}}( \mathcal{D}_{\mu_{f}} )$,
\begin{align}
\Lambda_{\mu_{f}}^{\prime}( s ) = a
\quad \Longrightarrow \quad
\Lambda_{\mu_{f}}^{\ast}( a ) = a \, s - \Lambda_{\mu_{f}}( s ) .
\label{eq:rate-function}
\end{align}

Similar to \sectref{sect:dms}, one can consider the notion of $\mu_{f}$ being a lattice measure.
We say that $\mu_{f}$ is a \emph{lattice measure} if $\mu_{f}$ is discrete%
\footnote{A measure $\nu$ is said to be \emph{discrete} if there exists a measurable set $\mathcal{E}$ such that it is countable and $\nu( \mathcal{E}^{\complement} ) = 0$, where $\mathcal{E}^{\complement}$ denotes the complement of $\mathcal{E}$.}
and there exists a positive constant $d$ such that $f( t_{1} ) - f( t_{2} )$ is a multiple of $d$ whenever $\mu_{f}( t_{1} ) \mu_{f}( t_{2} ) > 0$.
Otherwise, we say that $\mu_{f}$ is a \emph{nonlattice measure.}
For a lattice measure $\mu_{f}$, its \emph{maximal span} $d_{f}$ is defined by the maximum of those $d$.
For convenience, we set $d_{f} = 0$ if $\mu_{f}$ is nonlattice.
Then, given a positive parameter $s$, define
\begin{align}
\upsilon_{s}( f )
\coloneqq
\begin{dcases}
\frac{ d_{f} }{ \mathrm{e}^{d_{f} s} - 1 }
& \mathrm{if} \ \text{$\mu_{f}$ is lattice} ,
\\
s^{-1}
& \mathrm{if} \ \text{$\mu_{f}$ is nonlattice} .
\end{dcases}
\end{align}

Now, consider $n$ Borel-measurable functions $f_{1}, \dots, f_{n}$ in which $\mu \circ (f_{1}, \dots, f_{n})^{-1} = \mu_{f_{1}} \times \dots \times \mu_{f_{n}}$ and $\mu_{f_{i}} = \mu_{f}$ for each $1 \le i \le n$.%
\footnote{When $\mu$ is a probability measure, this implies that $f_{1}, \dots, f_{n}$ are i.i.d.\ copies of a real-valued r.v.\ $f$.}
The following lemma states a strong large deviations result known as the Bahadur--Rao theorem.

\begin{lemma}[{\cite[Theorem~3.7.4]{dembo_zeitouni_1998} and \cite[Chapter~VIII.4]{petrov_1975} for probability measures $\mu$}]
\label{lem:BR}
Let $a = \Lambda_{\mu_{f}}^{\prime}( s )$ for some positive $s \in \operatorname{\mathrm{int}}( \mathcal{D}_{\mu_{f}} )$.
Then, it holds that%
\footnote{When $\mu_{f}$ is lattice, then the remainder terms $+\mathrm{o}(1)$ can be refined as $+\mathrm{O}( n^{-1} )$; see \cite[Chapter~VIII.4]{petrov_1975}.}
\begin{align}
\mu \left\{ \sum_{i = 1}^{n} f_{i} > a \, n \right\}
& =
\frac{ \mathrm{e}^{ - n \Lambda_{\mu_{f}}^{\ast}( a ) } }{ \sqrt{ 2 \pi n \, \Lambda_{\mu_{f}}\mydprime( s ) } } \Big( \upsilon_{s}(f) + \mathrm{o}( 1 ) \Big) ,
\\
\mu \left\{ \sum_{i = 1}^{n} f_{i} = a \, n \right\}
& =
\frac{ \mathrm{e}^{ - n \Lambda_{\mu_{f}}^{\ast}( a ) } }{ \sqrt{ 2 \pi n \, \Lambda_{\mu_{f}}\mydprime( s ) } }\Big( d_{f} + \mathrm{o}( 1 ) \Big).
\end{align}
\end{lemma}

In \cite[Theorem~3.7.4]{dembo_zeitouni_1998} and \cite[Chapter~VIII.4]{petrov_1975}, \lemref{lem:BR} is stated in the case when $\mu$ is a probability measure, and its proof can be readily extended to $\sigma$-finite measures $\mu$.
We give a proof sketch of \lemref{lem:BR} in \appref{app:BR}.

The following lemma is a variant of \lemref{lem:BR}.

\begin{lemma}
\label{lem:SLD}
Let $\{ \mathcal{I}_{n} \}_{n = 1}^{\infty}$ be a sequence of real intervals, and $\mathcal{I} = \bigcup_{n} \mathcal{I}_{n}$.
Consider a real function $a_{n}( \cdot )$ on $\mathcal{I}$ for each $n \in \mathbb{N}$.
Suppose that $a = \Lambda_{\mu_{f}}^{\prime}( s )$ for some positive $s \in \operatorname{\mathrm{int}}( \mathcal{D}_{\mu_{f}} )$.
If
\begin{align}
a_{n}( t )
=
\mathrm{o}( n )
\label{eq:remainder_assumpsion}
\end{align}
uniformly on $\mathcal{I}_{n}$ as $n \to \infty$,
then there exist $r_{f}^{(1)}(n, t) = \mathrm{o}( 1 )$ and $r_{f}^{(2)}(n, t) = \mathrm{O}( 1 )$ uniformly on $\mathcal{I}_{n}$ as $n \to \infty$ such that
\begin{align}
&\mu \left\{ \sum_{i = 1}^{n} f_{i} > a \, n + a_{n}( t ) \right\}\nonumber\\*
&\qquad =
\frac{ \mathrm{e}^{ - K_{f}(n, s, t) } }{ \sqrt{ 2 \pi n \, \Lambda_{\mu_{f}}\mydprime( s ) \, (1 + \mathrm{o}(1)) } } \Big( \upsilon_{s}(f) + \mathrm{o}( 1 ) \Big) ,
\label{eq:SLD_1} \\
&\mu \left\{ \sum_{i = 1}^{n} f_{i} = a \, n + a_{n}( t ) \right\}\nonumber\\*
& \qquad=
\frac{ \mathrm{e}^{ - K_{f}(n, s, t) } }{ \sqrt{ 2 \pi n \, \Lambda_{\mu_{f}}\mydprime( s ) \, (1 + \mathrm{o}(1)) } }\Big( d_{f} + \mathrm{o}( 1 ) \Big)
\label{eq:SLD_2}
\end{align}
uniformly on $\mathcal{I}_{n}$ as $n \to \infty$, where the exponent part $K_{f}(n, s, t)$ is given as
\begin{align}
K_{f}(n, s, t)
=
n   \Lambda_{\mu_{f}}^{\ast}( a ) + s \, a_{n}( t ) + \frac{ a_{n}( t )^{2} }{ 2 \, n \, \Lambda_{\mu_{f}}\mydprime( s ) } \, (1 \!+\! \mathrm{o}(1))
\label{eq:K_f}
\end{align}
uniformly on $\mathcal{I}_{n}$ as $n \to \infty$.
\end{lemma}

\begin{IEEEproof}[Proof of \lemref{lem:SLD}]
See \appref{app:SLD}.
\end{IEEEproof}

\begin{remark}
\lemref{lem:SLD} is a minor extension of Hayashi's technical result \cite[Lemma~3]{hayashi_2018}.
The main differences vis-\`{a}-vis \cite[Lemma~3]{hayashi_2018} is that the %$\mu$ is not only finite but $\sigma$-finite  and the 
asymptotic expansion in~\eqref{eq:K_f} is refined.
\end{remark}

\subsection{Proof of \lemref{lem:MD_block}} \label{sec:proofLem3}

Denote by
$
\iota_{n}( X )
\coloneqq
- \ln P_{X^{n}}( X^{n} )
$
the information density of $X^{n}$, where $\iota(X) \coloneqq \iota_{1}(X)$.
Consider a $\sigma$-finite measure $\mu$ in which $\mu_{X}$ is the counting measure on $\operatorname{supp}(X)$ and $\mu_{X}( \mathcal{X} \setminus \operatorname{supp}(X) ) = 0$.
Now, define $\nu_{X} \coloneqq \mu \circ (-\iota(X))^{-1}$.
Since $H_{\alpha}(X) < \infty$ for some $0 < \alpha < 1$, we observe that $s \mapsto \Lambda_{\nu_{X}}( s )$ is infinitely differentiable at $s = 1$.
Then, a direct calculation shows
\begin{align}
\Lambda_{\nu_{X}}^{\prime}( 1 )
& =
- (\ln 2) \, H(X),
\label{eq:Lambda-H} \\
\Lambda_{\nu_{X}}\mydprime( 1 )
& =
(\ln 2)^{2} \, V(X) .
\label{eq:Lambda-V}
\end{align}

Choose the nonnegative number $\eta_{n}(\varepsilon_{n}, X)$ so that
\begin{align}
F_{n}^{+}(\varepsilon_{n}, X)
\coloneqq
\mathbb{P}\{ \iota_{n}(X) \le \eta_{n}(\varepsilon_{n}, X) \}
& \ge
1 - \varepsilon_{n} ,
\label{eq:eta_1} \\*
F_{n}^{-}(\varepsilon_{n}, X)
\coloneqq
\mathbb{P}\{ \iota_{n}(X) < \eta_{n}(\varepsilon_{n}, X) \}
& <
1 - \varepsilon_{n} ,
\label{eq:eta_2}
\end{align}
respectively.
It follows from \eqref{eq:HT} that%
\footnote{This identity is a consequence of the Neyman--Pearson lemma.}
\begin{align}
&M^{\ast}(n, \varepsilon_{n})\nonumber\\*
&=
\mu\{ \iota_{n}( X ) < \eta_{n}(\varepsilon_{n}, X) \}  \nonumber\\*
&\quad+ \left\lceil \left( \frac{ (1 - \varepsilon_{n}) - F_{n}^{-}(\varepsilon_{n}, X) }{ F_{n}^{+}(\varepsilon_{n}, X) - F_{n}^{-}(\varepsilon_{n}, X) } \right) \mu\{ \iota_{n}( X ) = \eta_{n}(\varepsilon_{n}, X) \} \right\rceil ,
\label{eq:k-ast_mu}
\end{align}
yielding that
\begin{align}
\mu\{ \iota_{n}( X ) < \eta_{n}(\varepsilon_{n}, X) \}
<
M^{\ast}(n, \varepsilon_{n})
\le
\mu\{ \iota_{n}( X ) \le \eta_{n}(\varepsilon_{n}, X) \} .
\label{eq:loose_M-mu}
\end{align}

Fix a positive number $r$ arbitrarily.
Define
\begin{align}
\lambda_{n}(\varepsilon_{n}, X)
\coloneqq
\frac{ \eta_{n}(\varepsilon_{n}, X) - n \, (\ln 2) \, H(X) }{ (\ln 2) \sqrt{n \, V(X)} } .
\label{def:lambda}
\end{align}
By Taylor's theorem for $s \mapsto \Phi^{-1}( s )$ around $s = 1 - \varepsilon_{n}$, we observe that
\begin{align}
\lambda_{n}(\varepsilon_{n}, X)
& =
\Phi^{-1}( 1 - \varepsilon_{n} ) - \frac{ (1 - \varepsilon_{n}) - \Phi( \lambda_{n}(\varepsilon_{n}, X) ) }{ f_{\mathrm{G}}( s_{n}(\varepsilon_{n}, X) ) } ,
\label{eq:Taylor_lambda-n}
\end{align}
where $0 < s_{n}(\varepsilon_{n}, X) < 1$ is given by
\begin{align}
s_{n}(\varepsilon_{n}, X)
\coloneqq
\theta_{n}(\varepsilon_{n}, X) \, (1 - \varepsilon_{n}) + (1 - \theta_{n}(\varepsilon_{n}, X)) \, \Phi( \lambda_{n}(\varepsilon_{n}, X) )
\label{def:s_n}
\end{align}
for some $0 \le \theta_{n}(\varepsilon_{n}, X) \le 1$.
Substituting \eqref{eq:Taylor_lambda-n} into \eqref{def:lambda}, we see that
\begin{align}
&\frac{ \eta_{n}(\varepsilon_{n}, X) }{ \ln 2 }\nonumber\\*
& =
n \, H(X) - \sqrt{ n \, V(X) } \, \left( \Phi^{-1}(\varepsilon_{n}) + \frac{ (1 - \varepsilon_{n}) - \Phi( \lambda_{n}(\varepsilon_{n}, X) ) }{ f_{\mathrm{G}}( s_{n}(\varepsilon_{n}, X) ) } \right) .
\label{eq:Taylor_eta-n_Phi-lambda-n}
\end{align}
Now, suppose that \eqref{eq:eps_small_range} holds.
 Then, we see from \eqref{eq:MD_1} of \lemref{lem:MD} that %there exist positive constants $A_{0} = A_{0}(X, r)$ and $B_{0} = B_{0}(X, r)$ such that
\begin{align}
\!\!s_{n}( \varepsilon_{n}, X )
& \!\le\!
1 - \varepsilon_{n} \left( 1  \!+\!\frac{ \cX\, \Phi^{-1}(\varepsilon_n )^3}{\sqrt{n}}\! -\! \mathrm{O}\bigg(\frac{ \Phi^{-1}(\varepsilon_n )}{\sqrt{n}} \bigg) \right).
%\notag \\
%& =
%1 - \varepsilon_{n} + \varepsilon_{n} \left( 1 - \exp\left( - \frac{ A_{0} \, \varepsilon_{n} \, \Phi^{-1}( \varepsilon_{n} )^{2} }{ n } \right) \right) + \frac{ B_{0} \, f_{\mathrm{G}}( \varepsilon_{n} ) }{ \sqrt{ n } }
%\notag \\
%& \overset{\mathclap{\text{(b)}}}{\le}
%1 - \varepsilon_{n} + \frac{ A_{0} \, \varepsilon_{n}^{2} \, \Phi^{-1}( \varepsilon_{n} )^{2} }{ n } + \frac{ B_{0} \, f_{\mathrm{G}}( \varepsilon_{n} ) }{ \sqrt{ n } }
\label{eq:sn_UB}
\end{align}
for sufficiently large $n$.
%, where
%\begin{itemize}
%\item
%(a) follows from \eqref{eq:MD_1} of \lemref{lem:MD}, and
%\item
%(b) follows from the   fact that $1-\exp(-y)\le y$ for all $y\in\mathbb{R}$ (a consequence of the mean value theorem).
%\end{itemize}
Similarly, we get that %there exist positive constants $A_{1} = A_{1}(X, r)$ and $B_{1} = B_{1}(X, r)$ such that
\begin{align}
\!\!s_{n}( \varepsilon_{n}, X )
& \!\ge\!
1 \!-\! \varepsilon_{n}  \left( 1 \! -\!\frac{ \cX  \Phi^{-1}(\varepsilon_n )^3}{\sqrt{n}} \!+\!\mathrm{O}\bigg(\frac{ \Phi^{-1}(\varepsilon_n )}{\sqrt{n}} \bigg) \right).
%s_{n}( \varepsilon_{n}, X )
%& \ge
%1 - \varepsilon_{n} - \frac{ A_{1} \, \varepsilon_{n}^{2} \, \Phi^{-1}( \varepsilon_{n} )^{2} }{ n } \, \exp\left( \frac{ A_{1} \, \varepsilon_{n} \, \Phi^{-1}( \varepsilon_{n} )^{2} }{ n } \right) - \frac{ B_{1} \, f_{\mathrm{G}}( \varepsilon_{n} ) }{ \sqrt{ n } }
\label{eq:sn_LB}
\end{align}
for sufficiently large $n$. It follows from \eqref{eq:inv-Gauss_1st}, \eqref{eq:inv-Gauss_f}, and \eqref{eq:eps_small_range} that 
\begin{equation}
\left| \frac{ \cX\, \Phi^{-1}(\varepsilon_n )^3}{\sqrt{n}} -\mathrm{O}\bigg( \frac{\Phi^{-1}(\varepsilon_n )}{\sqrt{n}} \bigg)   \right|\le  \mathrm{O} \left( \sqrt{\frac{\ln^3 n}{n} } \right).
\end{equation}
as $n \to \infty$. % for any constants $A$ and $B$.
Inserting this estimate into \eqref{eq:sn_UB}  and \eqref{eq:sn_LB}, we obtain
\begin{align}
s_{n}(\varepsilon_{n}, X)
=
1 - \varepsilon_{n} \left( 1 + \mathrm{O}\left( \sqrt{ \frac{ \ln^3 n }{ n } } \right) \right)
\end{align}
as $n \to \infty$.
Therefore, it follows from \eqref{eq:inv-Gauss_f} and Taylor's theorem for $s \mapsto f_{\mathrm{G}}( s )$ around $s = 1 - \varepsilon_{n}$ that there exists some $a_n $ between $1-\varepsilon_n$ and $1-\varepsilon_n+ \mathrm{O}( ( n^{-1} \ln n )^{1/2} )$ such that 
\begin{align}
f_{\mathrm{G}}( s_{n}( \varepsilon_{n}, X ) )
& =
f_{\mathrm{G}}( 1-\varepsilon_n) + \mathrm{O}\left( f_{\mathrm{G}}^{\prime}(a_{n}) \, \varepsilon_{n}  \sqrt{  \frac{\ln^3 n}{n} } \right)
\notag \\
&= f_{\mathrm{G}}( \varepsilon_n) + \mathrm{O}\left( \Phi^{-1}(a_{n}) \, \varepsilon_{n}  \sqrt{  \frac{\ln^3 n}{n} } \right)
\notag \\
& =
f_{\mathrm{G}}( \varepsilon_{n} ) + \mathrm{O}\left( \frac{ \varepsilon_{n} \ln^2 n }{ \sqrt{ n } } \right) \notag \\
&= f_{\mathrm{G}}( \varepsilon_{n} )(1+ \mathrm{o}(1))
\label{eq:Gauss_f_sn}
\end{align}
as $n \to \infty$. On the other hand, from \eqref{eq:MD_1} of \lemref{lem:MD},   %there exist positive constants $A_{2} = A_{2}(X, r)$ and $B_{2} = B_{2}(X, r)$ such that
\begin{align}
\Big| (1\! -\! \varepsilon_{n}) \!-\!  \Phi ( \lambda_{n}(\varepsilon_{n}, X) ) \Big|
& \! \le \! \varepsilon_n \left|\frac{ \cX\, \Phi^{-1}(\varepsilon_n )^3}{\sqrt{n}} \!-\!\mathrm{O}\bigg( \frac{\Phi^{-1}(\varepsilon_n )}{\sqrt{n}} \bigg)    \right|.
\label{eq:abs-gap_eps-invGauss}
\end{align}
for sufficiently large $n$.
%, where
%\begin{itemize}
%\item
%(a) follows from \eqref{eq:MD_1} of \lemref{lem:MD}, and
%\item
%(b) follows from the fact that  $\exp(y)-1\le y\exp( y)$ for all $y\in\mathbb{R}$ (a consequence of the mean value theorem).
%\end{itemize}
Combining \eqref{eq:Gauss_f_sn} and \eqref{eq:abs-gap_eps-invGauss}, we have from \eqref{eq:inv-Gauss_1st} and \eqref{eq:inv-Gauss_f} that
\begin{align}
 \left|\frac{ (1 - \varepsilon_{n}) - \Phi^{-1}( \lambda_{n}(\varepsilon_{n}, X) ) }{ f_{\mathrm{G}}( s_{n}( \varepsilon_{n}, X ) ) } \right|
%& \le
%\frac{ \varepsilon_n \left|\frac{ \cX\, \Phi^{-1}(\varepsilon_n )^3}{\sqrt{n}} -\mathrm{O}\bigg( \frac{\Phi^{-1}(\varepsilon_n )}{\sqrt{n}} \bigg)    \right|  
%}{\left| f_{\mathrm{G}}( \varepsilon_{n} ) + \mathrm{O}\left( \dfrac{ \varepsilon_{n} \ln^2 n }{ \sqrt{ n } } \right)\right|}   
%\notag \\
& \le
\varepsilon_n \left|\frac{  \frac{ \cX\, \Phi^{-1}(\varepsilon_n )^3}{\sqrt{n}} -\mathrm{O}\bigg( \frac{\Phi^{-1}(\varepsilon_n )}{\sqrt{n}} \bigg)     }{  f_{\mathrm{G}}( \varepsilon_{n} )  (1+\mathrm{o}(1))} \right|
\notag \\
& =\frac{1}{\sqrt{n}} \left(  \cX\, \Phi^{-1}(\varepsilon_n)^2+\mathrm{O}(1) \right),
%\mathrm{O}\left( \frac{ 1 }{ \sqrt{n} } \right) 
\label{eq:lambda-n_MD}
\end{align}
as $n \to \infty$.
Hence, it follows from \eqref{eq:Taylor_eta-n_Phi-lambda-n} and \eqref{eq:lambda-n_MD} that
\begin{align}
\frac{ \eta_{n}(\varepsilon_{n}, X) }{ \ln 2 }
& =
n \, H(X) - \sqrt{ n \, V(X) } \, \Phi^{-1}(\varepsilon_{n}) \nonumber\\*
&\qquad+ \left( \cX\,  \Phi^{-1}(\varepsilon_n)^2 +\mathrm{O}( 1 )\right)
\label{eq:eta_MD}
\end{align}
as $n \to \infty$, provided that \eqref{eq:eps_small_range} holds.
By using \eqref{eq:MD_2} of \lemref{lem:MD} rather than \eqref{eq:MD_1}, we can prove \eqref{eq:eta_MD} in an analogous manner even if the sequence $\{ \varepsilon_{n} \}_{n = 1}^{\infty}$ satisfies that
\begin{align}
\frac{ 1 }{ 2 }
\le
\varepsilon_{n}
\le
1 - \frac{ 1 }{ n^{r} } .
\end{align}
Therefore, we conclude that \eqref{eq:eta_MD} holds for every sequence $\{ \varepsilon_{n} \}_{n = 1}^{\infty}$ satisfying \eqref{eq:polynomial_eps}.
Now, note from \eqref{eq:Lambda-H} and \eqref{eq:Lambda-V} that \eqref{eq:eta_MD} can be rewritten as
\begin{align}
\eta_{n}(\varepsilon_{n}, X)
& =
- n \, \Lambda_{\nu_{X}}^{\prime}( 1 ) - \sqrt{ n \, \Lambda_{\nu_{X}}\mydprime( 1 ) } \, \Phi^{-1}(\varepsilon_{n}) \nonumber\\* 
&\qquad+\left( \Phi^{-1}(\varepsilon_n)^2\, \cX +\mathrm{O}( 1 )\right)
\label{eq:eta_MD2}
\end{align}
as $n \to \infty$.
Therefore, applying \eqref{eq:SLD_1} of \lemref{lem:SLD} with
\begin{align}
s
& =
1 ,
\\
f_{i}
& =
\ln P_{X_{i}}( X_{i} )
\qquad (\mathrm{for} \ i = 1, \dots, n) ,
\\
a_{n}( \varepsilon_{n} )
& =
\sqrt{ n \, \Lambda_{\nu_{X}}\mydprime( 1 ) } \, \Phi^{-1}(\varepsilon_{n}) + \mathrm{O}( 1 )
\end{align}
we obtain~\eqref{eq:SLD_mu_1}  (at the top of the next page)
\begin{figure*}
\begin{align}
&
\mu\{ \iota_{n}(X) < \eta_{n}(\varepsilon_{n}, X) \}
\notag \\
& =
\mu\left\{ \sum_{i = 1}^{n} \ln P_{X_{i}}(X_{i}) > n \, \Lambda_{\nu_{X}}^{\prime}( 1 ) + \sqrt{ n \, \Lambda_{\nu_{X}}\mydprime( 1 ) } \, \Phi^{-1}(\varepsilon_{n}) + \mathrm{O}( 1 ) \right\}
\notag \\
& =
\frac{ \upsilon(X) + \mathrm{o}( 1 ) }{ \sqrt{2 \pi n \, \Lambda_{\nu_{X}}\mydprime( 1 ) \, (1 + \mathrm{o}(1)) } } \exp\bigg( - n \, \Lambda_{\nu_{X}}^{\ast}( \Lambda_{\nu_{X}}^{\prime}( 1 ) ) - \left( \sqrt{ n \, \Lambda_{\nu_{X}}\mydprime( 1 ) } \, \Phi^{-1}(\varepsilon_{n}) + \mathrm{O}( 1 ) \right)
\notag \\
&\qquad \qquad \qquad \qquad \qquad \qquad \qquad \qquad {}
- \frac{ 1 }{ 2 \, n \, \Lambda_{\nu_{X}}\mydprime( 1 ) } \left( \sqrt{ n \, \Lambda_{\nu_{X}}\mydprime( 1 ) } \, \Phi^{-1}(\varepsilon_{n}) + \mathrm{O}( 1 ) \right)^{2} \bigg)
\notag \\
& =
\frac{ \upsilon(X) + \mathrm{o}( 1 ) }{ \sqrt{2 \pi n \, (\ln 2)^{2} \, V(X) \, (1 + \mathrm{o}(1)) } } \exp\left( n \, (\ln 2) \, H(X) - (\ln 2) \sqrt{ n \, V(X) } \, \Phi^{-1}(\varepsilon_{n}) + \left( \cX-\frac{1}{2} \right)  \Phi^{-1}( \varepsilon_{n} )^{2}   + \mathrm{O}( 1 ) \right)
\notag \\
& =
\frac{ \upsilon(X) + \mathrm{o}( 1 ) }{ \sqrt{2 \pi n \, (\ln 2)^{2} \, V(X) \, (1 + \mathrm{o}(1)) } } \exp\left( \eta_{n}(\varepsilon_{n}, X) +\left( \, \cX-\frac{1}{2} \right) \Phi^{-1}( \varepsilon_{n} )^{2}  +   \mathrm{O}( 1 )\right)
\label{eq:SLD_mu_1}
\end{align}
\hrulefill
\end{figure*}
as $n \to \infty$, where $\upsilon(X)$ is given as
\begin{align}
\upsilon(X)
\coloneqq
\begin{dcases}
\frac{ (\ln 2) \, d_{X} }{ 2^{d_{X}} - 1 }
& \mathrm{if} \ X \ \mathrm{is} \ \mathrm{a} \ \mathrm{lattice} \ \mathrm{source} ,
\\
1
& \mathrm{if} \ X \ \mathrm{is} \ \mathrm{a} \ \mathrm{nonlattice} \ \mathrm{source} ,
\end{dcases}
\end{align}
and $d_{X}$ is defined in \sectref{sect:dms}.
Analogously, we get from \eqref{eq:SLD_2} of \lemref{lem:SLD} that
\begin{align}
&\mu\{ \iota_{n}(X) = \eta_{n}(\varepsilon_{n}, X) \} \nonumber\\*
&=
\frac{ (\ln 2) \, d_{X} + \mathrm{o}( 1 ) }{ \sqrt{2 \pi n \, (\ln 2)^{2} \, V(X) \, (1 + \mathrm{o}(1)) } }\nonumber\\*
&\quad\times \exp\left( \eta_{n}(\varepsilon_{n}, X)  +\left(  \cX-\frac{1}{2} \right) \Phi^{-1}( \varepsilon_{n} )^{2}   + \mathrm{O}( 1 ) \right)
\label{eq:SLD_mu_2}
\end{align}
as $n \to \infty$.
Combining \eqref{eq:loose_M-mu}, \eqref{eq:SLD_mu_1}, and \eqref{eq:SLD_mu_2}, we obtain
\begin{align}
&\ln M^{\ast}(n, \varepsilon_{n})\nonumber\\*
& =
\eta_{n}(\varepsilon_{n}, X) + \left(  \cX-\frac{1}{2} \right) \Phi^{-1}( \varepsilon_{n} )^{2}  \nonumber\\*
&\qquad - \frac{ 1 }{ 2 } \ln \Big( 2 \pi n \, (\ln 2)^{2} \, V(X) \, (1 + \mathrm{o}(1)) \Big) + \mathrm{O}( 1 )
\notag \\
& =
\eta_{n}(\varepsilon_{n}, X) - \frac{ 1 }{ 2 } \ln n +\left(  \cX-\frac{1}{2} \right) \Phi^{-1}( \varepsilon_{n} )^{2}  + \mathrm{O}( 1 )
\end{align}
as $n \to \infty$.
This completes the proof of \lemref{lem:MD_block}.
\hfill\IEEEQEDhere

\section{Concluding Remarks and Future Works}
\label{sect:conclusion}

 In this study, we investigated the third-order asymptotics of the problem of variable-length compression allowing errors. Our main contribution is in refining the second-order asymptotic expansion of  Kostina, Polyanskiy, and Verd\'{u} \cite{kostina_polyanskiy_verdu_2015} to obtain the third-order term which is $-(1-\varepsilon)(\log n)/2$ where $\varepsilon$ is the permissible error probability in reconstructing the source. Our proof strategy demonstrates   a novel utility of a combination of moderate deviations  (or Cram\'er-type large deviations) and strong large deviations analyses in information theory.

 One extension of the work herein is to consider the third-order term in the classical channel coding problem~\cite{polyanskiy_poor_verdu_2010,tan_tomamichel_2015,tomamichel_tan_2013}   in the moderate deviations regime, i.e., the analogue of \lemref{lem:MD_block} for the channel coding setting.  By evaluating Polyanskiy, Poor, Verd\'u's random coding union (RCU) bound~\cite[Theorem~16]{polyanskiy_poor_verdu_2010} and the meta-converse~\cite[Theorem~27]{polyanskiy_poor_verdu_2010} with carefully chosen output distributions (e.g., in \cite{tomamichel_tan_2013}), and replacing the use of the Berry-Esseen theorem with Lemma~\ref{lem:refined_MD} and~\ref{lem:MD} of the present paper, this should yield the third-order term in the moderate deviations regime. Such a strategy may also be amenable to additive white Gaussian noise (AWGN) channels \cite{tan_tomamichel_2015}.

%Our main result is stated in \thref{th:third} of \sectref{sect:main}, which shows that the first-, second-, and third-order coding rates depend on the permissible probability of error $0 \le \varepsilon \le 1$.
%This observation differs from the third-order asymptotics of the fixed-length compression problem as stated in \eqref{eq:third_fixed}, which shows that the first- and third-order coding rates do not depend on $\varepsilon$.
%To derive \thref{th:third}, we leveraged certain \emph{moderate deviations} and \emph{strong large deviations} results for the fixed-length compression problem in \lemref{lem:MD_block} of \sectref{sect:fixed}.
%This proof strategy, together with a connection between the variable- and fixed-length compression problems (see \sectref{sect:cutoff-log}), shows a novel utility of a combination of moderate deviations and strong large deviations analyses in information theory.
%Since the fundamental limits of fixed-length compression problems are intimately related to those of variable-length compression problems under excess length constraints (cf.\ \cite{merhav_1991, kontoyiannis_verdu_2014, kosut_sankar_2017, saito_matsushima_2017, iri_kosut_2019, nomura_yagi_2019}), similar to \lemref{lem:MD_block}, moderate deviations and strong large deviations analyses of the latter class of problems might be of interest in future work.
%Finally, we believe that the mathematical results in  \lemref{lem:MD} and \lemref{lem:SLD} may be of independent interest.

\appendices

\section{Proof of \lemref{lem:cutoff-log_integral-M}}
\label{app:cutoff-log_integral-M}

To prove \lemref{lem:cutoff-log_integral-M}, we first choose a positive integer $\xi_{n} = \xi_{n}(n, X)$ so that%
\footnote{Note that $\xi_{n}$ plays the role of $\tilde{\xi}_{n}$ defined in \eqref{eq:xi_tilde_1}--\eqref{eq:xi_tilde_2}.
In fact, it is clear that $\xi_{n} = \log \tilde{\xi}_{n}$.}
\begin{align}
\mathbb{P}\{ \gamma_n ( X^{n} ) \ge \xi_{n} \}
& \ge
\varepsilon ,
\label{eq:eta-n_1} \\
\mathbb{P}\{ \gamma_n ( X^{n} ) > \xi_{n} \}
& <
\varepsilon
\label{eq:eta-n_2}
\end{align}
for each $0 < \varepsilon \le 1$, and $\xi_{n} = 2^{n H_{0}(X)}$ if $\varepsilon = 0$, where
\begin{align}
H_{0}(X)
\coloneqq
\begin{cases}
\log |\operatorname{supp}(X)|
& \mathrm{if} \ \operatorname{supp}(X) \ \mathrm{is} \ \mathrm{finite} ,
\\
\infty
& \mathrm{if} \ \operatorname{supp}(X) \ \mathrm{is} \ \mathrm{infinite} .
\end{cases}
\end{align}
It is clear that $\xi_{n} = 2^{n H_{\infty}(X)}$ if $\varepsilon = 1$, where $H_{\infty}(X) \coloneqq - \log \max_{x \in \mathcal{X}} P_{X}( x )$ denotes the min-entropy.
Since $\mathbb{E}[ \langle \log \gamma_n ( X^{n} ) \rangle_{\varepsilon} ] = 0$ if $\varepsilon \ge 1 - 2^{- n H_{\infty}(X)}$, it suffices to consider the case when $0 \le \varepsilon < 1 - 2^{- n H_{\infty}(X)}$.
Then, a direct calculation shows that
\begin{align}
&\mathbb{E}[ \langle \log \gamma_n ( X^{n} ) \rangle_{\varepsilon} ]\nonumber\\*
& \overset{\mathclap{\text{(a)}}}{=}
\int_{0}^{\infty} \mathbb{P}\{ \langle \log \gamma_n ( X^{n} ) \rangle_{\varepsilon} > s \} \, \mathrm{d} s
 \\
& \overset{\mathclap{\text{(b)}}}{\le}
\int_{0}^{\log \xi_{n}} \mathbb{P}\{ s < \log \gamma_n ( X^{n} ) \le \log \xi_{n} \} \, \mathrm{d} s
 \\
& \overset{\mathclap{\text{(c)}}}{=}
\int_{0}^{\infty} \mathbb{P}\{ \log \gamma_n ( X^{n} ) > s \} \, \mathrm{d} s - \int_{\log \xi_{n}}^{\infty} \mathbb{P}\{ \log \gamma_n ( X^{n} ) > t \} \, \mathrm{d} t\nonumber\\*
&\qquad{} - (\log \xi_{n}) \, \mathbb{P}\{ \gamma_n ( X^{n} ) > \xi_{n} \}
 \\
& =
\int_{0}^{\infty} \mathbb{P}\{ \log \gamma_n ( X^{n} ) > s \} \, \mathrm{d} s - \int_{\log \xi_{n}}^{\log (\xi_{n} + 1)} \mathbb{P}\{ \log \gamma_n ( X^{n} ) > t \} \, \mathrm{d} t 
\notag \\
& \qquad {}
- \sum_{k = \xi_{n} + 1}^{\infty} \int_{\log k}^{\log(k+1)} \mathbb{P}\{ \log \gamma_n ( X^{n} ) > u \} \, \mathrm{d} u \nonumber\\*
&\qquad{ } - (\log \xi_{n}) \, \mathbb{P}\{ \gamma_n ( X^{n} ) > \xi_{n} \}
 \\
& =
\int_{0}^{\infty} \mathbb{P}\{ \log \gamma_n ( X^{n} ) > s \} \, \mathrm{d} s \nonumber\\*
&\qquad { }- (\log (\xi_{n} + 1) - \log \xi_{n}) \, \mathbb{P}\{ \log \gamma_n ( X^{n} ) > \log \xi_{n} \}
\notag \\
& \qquad {}
- \sum_{k = \xi_{n} + 1}^{\infty} (\log (k+1) - \log k) \, \mathbb{P}\{ \log \gamma_n ( X^{n} ) > \log k \} \nonumber\\*
& \qquad {}- (\log \xi_{n}) \, \mathbb{P}\{ \gamma_n ( X^{n} ) > \xi_{n} \}
 \\
& =
\int_{0}^{\infty} \mathbb{P}\{ \log \gamma_n ( X^{n} )\! >\! s \} \, \mathrm{d} s\! - \!\sum_{k = 1 + \xi_{n}}^{\infty} (\log k) \, \mathbb{P}\{ \gamma_n ( X^{n} ) = k \}
 \\
& =
\sum_{k = 2}^{\xi_{n}} (\log k) \, \mathbb{P}\{ \gamma_n ( X^{n} ) = k \}
 \\
& =
\sum_{k = 2}^{\xi_{n}} (\log k) \, \int_{\mathbb{P}\{ \gamma_n ( X^{n} ) > k \}}^{\mathbb{P}\{ \gamma_n ( X^{n}) \ge k \}} \mathrm{d} s
  \\
& \overset{\mathclap{\text{(d)}}}{=}
\int_{\mathbb{P}\{ \gamma_n ( X^{n} ) > \xi_{n} \}}^{1 - 2^{-n H_{\infty}(X)}} \log (M^{\ast}(n, s) +1)\, \mathrm{d} s
 \\
& =
\int_{\varepsilon}^{1 - 2^{-n H_{\infty}(X)}} \log(M^{\ast}(n, s) +1) \, \mathrm{d} s \nonumber\\* 
&\qquad {} + (\varepsilon - \mathbb{P}\{ \gamma_n ( X^{n} ) > \xi_{n} \}) \log (M^{\ast}(n, \varepsilon)+1)
 \\
& \overset{\mathclap{\text{(e)}}}{\le}
\int_{\varepsilon}^{1 - 2^{-n H_{\infty}(X)}} \log(M^{\ast}(n, s) +1) \, \mathrm{d} s  \nonumber\\* 
&\qquad {}  \mathbb{P}\{ \gamma_n ( X^{n} ) = \xi_{n} \} \log (M^{\ast}(n, \varepsilon)+1)
  \\
& \overset{\mathclap{\text{(f)}}}{\le}
\int_{\varepsilon}^{1 - 2^{-n H_{\infty}(X)}} \log(M^{\ast}(n, s) +1) \, \mathrm{d} s  \nonumber\\* 
&\qquad {} + 2^{- n H_{\infty}(X)} \log (M^{\ast}(n, \varepsilon)+1) ,
\label{eq:chain1}
\end{align}
where
\begin{itemize}
\item
(a) follows from the fact that $
\mathbb{E}[ Z ]
=
\int_{0}^{\infty} \mathbb{P}\{ Z > z \} \, \mathrm{d} z
$
for every nonnegative-real-valued r.v.\ $Z$,
\item
(b) follows from \eqref{eq:eta-n_2},
\item 
 (c) follows    by the following elementary calculation
\begin{align}  
&\int_0^a\mathbb{P}\{s< Y\le a\}\, \mathrm{d}s + a\mathbb{P}\{ Y>a\} + \int_a^\infty \mathbb{P}\{ Y>s\}\, \mathrm{d}s  \notag\\
&=\int_0^a\mathbb{P}\{s\!<\! Y\!\le\! a\}\, \mathrm{d}s \! + \!\int_0^a \mathbb{P}\{ Y\!>\! a\} \, \mathrm{d}s\!+ \!\int_a^\infty \mathbb{P}\{ Y\!>\! s\}\, \mathrm{d}s  \notag\\
&=\int_0^a\mathbb{P}\{Y\!>\!s\}\, \mathrm{d}s   + \int_a^\infty \mathbb{P}\{ Y\!>\!s\}\, \mathrm{d}s \!= \!\int_0^\infty\mathbb{P}\{Y\!>\! s\}\, \mathrm{d}s    
\end{align}
and taking $Y=\log \gamma_n ( X^{n} )$ and $a = \log \xi_{n}$,
\item
(d) follows from the fact that $M^{\ast}(n, s) = k - 1$ if $\mathbb{P}\{ \gamma_n ( X^{n} ) > k \} < s < \mathbb{P}\{ \gamma_n ( X^{n} ) \ge k \}$,
\item
(e) follows from \eqref{eq:eta-n_1}, and
\item
(f) follows from the fact that
\begin{align}
2^{- n H_{\infty}(X)} = \mathbb{P}\{ \gamma_n (X^{n}) = 1 \} \ge \mathbb{P}\{ \gamma_n (X^{n}) = k \}
\end{align}
for every integer $k \ge 1$.
\end{itemize}
Analogously, we see that
\begin{align}
&\mathbb{E}[ \langle \log \gamma_n ( X^{n} ) \rangle_{\varepsilon} ]\nonumber\\* 
&\quad
 \ge
\int_{\varepsilon}^{1} \log (M^{\ast}(n, s) -1)\, \mathrm{d} s - 2^{- n H_{\infty}(X)} \log (M^{\ast}(n, \varepsilon)-1)
\label{eq:chain2}
\end{align}
Combining \eqref{eq:chain1} and \eqref{eq:chain2} and using the fact that $\log M^*(n,s)$ is  $\Theta(n)$, we obtain \lemref{lem:cutoff-log_integral-M}.
\hfill\IEEEQEDhere

\section{Proof of \lemref{lem:MD}}
\label{app:MD}
For the sake of brevity, we use $\mathrm{Q}(z) := 1-\Phi(z)=\Phi(-z)$, the complementary Gaussian cumulative distribution function, in this proof. In the following, we only prove~\eqref{eq:MD_2}; the proof for \eqref{eq:MD_1} follows analogously.  Starting from~\eqref{eqn:FnZn} in Lemma~\ref{lem:refined_MD}, we have 
\begin{equation}
F_n(-z_n) = \mathrm{Q}( z_{n} )  \, \exp\left(- \frac{ z_{n}^{3} \, \cZ}{ \sqrt{n}   } \right) + \mathrm{O}\left(\frac{1}{\sqrt{n}} \exp\bigg( - \frac{z_n^2}{2}  \bigg) \right) , \label{eqn:defFn}
\end{equation}
where the final term  (involving $\mathrm{O}(\cdot)$) results from the definition of the Gaussian probability density function  $\varphi(u) = \frac{1}{\sqrt{2\pi}} \mathrm{e}^{-u^{2}/2}$. We set this to be equal to $\varepsilon_n$ to solve for $z_n$, i.e., 
\begin{equation}
  F_n(-z_n) =\varepsilon_n. \label{eqn:FnZn_1}
\end{equation}  
From the usual central limit theorem \eqref{eq:ordinary-CLT}, we deduce that $z_n  = \mathrm{Q}^{-1}( \varepsilon_n (1+\mathrm{o}(1)) )$. Thus,  we can parametrize it as 
\begin{equation}
z_n =  \mathrm{Q}^{-1}\big( \varepsilon_n \exp(g(\varepsilon_n)) - h(\varepsilon_n) \big), \label{eqn:zn_param}
\end{equation} 
for some functions  $g(\varepsilon_n)$ and $h(\varepsilon_n)$ that tend  to zero as $n\to\infty$ and $h(\varepsilon_n)\to 0$ faster than $\varepsilon_n$, i.e.,
\begin{equation}
g(\varepsilon_n)=\mathrm{o}(1)\quad\mbox{and}\quad   h(\varepsilon_n) = \mathrm{o}(\varepsilon_n). \label{eqn:gh}
\end{equation}
 With these constraints on $g(\varepsilon_n)$ and $h(\varepsilon_n)$, we see that $z_n$ as parametrized in \eqref{eqn:zn_param} indeed satisfies the condition that $z_n  = \mathrm{Q}^{-1}( \varepsilon_n (1+\mathrm{o}(1)) )$. Now, we substitute~\eqref{eqn:zn_param} into  \eqref{eqn:defFn} and \eqref{eqn:FnZn_1} to obtain
\begin{align}
&\big(\varepsilon_n \exp(g(\varepsilon_n)) - h(\varepsilon_n) \big)\nonumber\\*
&\times\exp\left(-  \frac{\cZ\,\mathrm{Q}^{-1}\big( \varepsilon_n \exp(g(\varepsilon_n)) - h(\varepsilon_n) \big)^3}{\sqrt{n}} \right) \nonumber\\*
&\qquad+ \mathrm{O}\left(\frac{1}{\sqrt{n}}\exp\bigg(- \frac{\mathrm{Q}^{-1}\big( \varepsilon_n \exp(g(\varepsilon_n)) - h(\varepsilon_n) \big)^2}{2}\bigg) \right)=\varepsilon_n . \label{eqn:set}
\end{align}
Solving for the function $g$   by equating coefficients in~\eqref{eqn:set}, we obtain
\begin{align}
g(\varepsilon_n) &=\frac{\cZ\,\mathrm{Q}^{-1}\big( \varepsilon_n \exp(g(\varepsilon_n)) - h(\varepsilon_n) \big)^3}{\sqrt{n}} (1+\mathrm{o}(1))\\
&=\frac{\cZ\,\mathrm{Q}^{-1}\big(  \varepsilon_n   \big)^3}{\sqrt{n}} (1+\mathrm{o}(1)) ,
\end{align}
where the last equality follows from the properties of $g$ and $h$ in~\eqref{eqn:gh}. 
Solving for the function $h$   by equating coefficients in~\eqref{eqn:set}, we obtain
\begin{align}
 h(\varepsilon_n) &= \mathrm{O}\Bigg(\frac{1}{\sqrt{n}}\exp\bigg(- \frac{\mathrm{Q}^{-1}\big( \varepsilon_n \exp(g(\varepsilon_n)) - h(\varepsilon_n) \big)^2}{2}\nonumber\\* 
&\qquad {} 
+ \frac{\cZ\,\mathrm{Q}^{-1}\big( \varepsilon_n \exp(g(\varepsilon_n)) - h(\varepsilon_n) \big)^3 }{\sqrt{n}} \bigg) \Bigg).
\end{align}
Since $z_n = \mathrm{O}(n^{1/6})$ the second term in the sum of the exponent is asymptotically negligible (compared to the first term in the exponent) and so 
\begin{align}
 h(\varepsilon_n) %&= \mathrm{O}\Bigg(\frac{1}{\sqrt{n}}\exp\bigg(- \frac{\mathrm{Q}^{-1}\big( \varepsilon_n \exp(g(\varepsilon_n)) - h(\varepsilon_n) \big)^2}{2} \Bigg) \\
 &= \mathrm{O}\Bigg(\frac{1}{\sqrt{n}}\exp\bigg(- \frac{\mathrm{Q}^{-1}\big( \varepsilon_n (1+\mathrm{o}(1))\big)^2}{2} \bigg)\Bigg) \nonumber\\* 
&  =\mathrm{O}\Bigg(\frac{\varepsilon_n}{\sqrt{n}} \mathrm{Q}^{-1}\big( \varepsilon_n \big) \Bigg),
\end{align}
where the final equality holds because $\mathrm{Q}^{-1}(s) \sim \sqrt{2 \ln \frac{1}{s}}$ as $s\to 0^+$; see \eqref{eq:inv-Gauss_1st}.  Note that both  the derived $g$ and $h$ satisfy the requirements in \eqref{eqn:gh}. Therefore, 
\begin{align}
z_n &\!=\! \mathrm{Q}^{-1}\left( \varepsilon_n\exp\bigg(\cZ\, \frac{\mathrm{Q}^{-1}\big( \varepsilon_n   \big)^3}{\sqrt{n}} (1\!+\!\mathrm{o}(1))\bigg) \!-\! \mathrm{O}\Bigg(\frac{\varepsilon_n}{\sqrt{n}} \mathrm{Q}^{-1}\big( \varepsilon_n \big) \Bigg) \right) \nonumber\\*
&\!=\!\mathrm{Q}^{-1}\left( \varepsilon_n \bigg(1+ \frac{\cZ\,\mathrm{Q}^{-1}\big( \varepsilon_n   \big)^3}{\sqrt{n}}   - \mathrm{O}\Bigg(\frac{\mathrm{Q}^{-1}\big( \varepsilon_n \big) }{\sqrt{n}} \Bigg) \right), \label{eqn:taylor_exp}
\end{align}
where \eqref{eqn:taylor_exp} follows by the fact that $\exp(x) = 1+x+\mathrm{o}(x)$ as $x\to 0$. 
Hence, we have 
\begin{align}
&\varepsilon_n \left( 1 -\frac{\cZ\,\Phi^{-1} (\varepsilon_n)^3 }{\sqrt{n}}-\mathrm{O}\bigg( \frac{\Phi^{-1} (\varepsilon_n)}{\sqrt{n}} \bigg) \right) =\mathrm{Q}(z_n) \nonumber\\* 
&\qquad {} = \Phi(-z_n) =  \Phi(\zeta_n(1-\varepsilon_n)) 
\end{align}
completing the proof of~\eqref{eq:MD_2}.
\hfill\IEEEQEDhere

\section{Proof of \lemref{lem:BR}}
\label{app:BR}

After some algebra, we get
\begin{align}
\mu \left\{ \sum_{i = 1}^{n} f_{i} \ge a \right\}
=
\mathrm{e}^{- n \Lambda_{\mu_{f}}^{\ast}(a)} \int_{0}^{\infty} \mathrm{e}^{-s t \sqrt{ n \Lambda_{\mu_{f}}\mydprime(s) }} \, \mathrm{d} F_{n}( t ) ,
\label{eq:tilde-mu}
\end{align}
where $F_{n}$ is a distribution function of the r.v.\ $W_{n}$ given by
\begin{align}
W_{n}
=
\frac{ 1 }{ \sqrt{ n \, \Lambda_{\mu_{f}}\mydprime( s ) } } \sum_{i = 1}^{n} (U_{i} - a) ,
\end{align}
and $U_{1}, \dots, U_{n}$ are i.i.d.\ r.v.'s with generic distribution $\tilde{\mu}_{f}$ constructed by the Radon--Nikodym derivative
\begin{align}
\frac{ \mathrm{d} \tilde{\mu}_{f} }{ \mathrm{d} \mu_{f} }( t )
=
\mathrm{e}^{s t - \Lambda_{\mu_{f}}( s )} .
\end{align}
Then, \lemref{lem:BR} can be proven by applying the Edgeworth expansion (cf.\ \cite[Chapter~XVI.4]{feller_1971} or \cite[Chapter~VI.3]{petrov_1975}) to the distribution $F_{n}$ in \eqref{eq:tilde-mu}; see the proof of \cite[Theorem~3.7.4]{dembo_zeitouni_1998}.
\hfill\IEEEQEDhere

\section{Proof of \lemref{lem:SLD}}
\label{app:SLD}

For the sake of brevity, we write
\begin{align}
\alpha_{n}
=
\alpha_{n}( t )
\coloneqq
a \, n + a_{n}( t ) .
\label{eq:asympt-alpha_n}
\end{align}
As in \cite[Exercise~2.2.24]{dembo_zeitouni_1998}, we observe that $\Lambda_{\mu_{f}}^{\ast}( \tilde{a} )$ is of class $C^{\infty}$ in $\tilde{a} \in \operatorname{\mathrm{int}}( \mathcal{F}_{\mu_{f}} )$, where $\mathcal{F}_{\mu_{f}} \coloneqq \{ \Lambda_{\mu_{f}}^{\prime}( \tilde{s} ) \mid \tilde{s} \in \operatorname{\mathrm{int}}( \mathcal{D}_{\mu_{f}} ) \}$.
Thus, since $s \in \operatorname{\mathrm{int}}( \mathcal{D}_{\mu_{f}} )$, it follows from \eqref{eq:rate-function}, \eqref{eq:remainder_assumpsion}, and \eqref{eq:asympt-alpha_n} that $\alpha_{n}/n \in \operatorname{\mathrm{int}}( \mathcal{F}_{\mu_{f}} )$ for sufficiently large $n$.
Henceforth, we assume that $n$ is large enough.
Noting this fact, denote by $s_{n} = s_{n}( t )$ the root of the equation $n \, \Lambda_{\mu_{f}}^{\prime}( \tilde{s} ) = \alpha_{n}$ with respect to $\tilde{s} \in \operatorname{\mathrm{int}}( \mathcal{D}_{\mu_{f}} )$.

Since $\Lambda_{\mu_{f}}^{\prime}( \cdot )$ is of class $C^{\infty}$ on $\operatorname{\mathrm{int}}( \mathcal{D}_{\mu_{f}} )$, and since \eqref{eq:remainder_assumpsion} implies that $\alpha_{n} \to a$ uniformly on $\mathcal{I}_{n}$ as $n \to \infty$, we see that $s_{n} \to s$ as $n \to \infty$ uniformly on $\mathcal{I}_{n}$.
In addition, it follows from Taylor's theorem for $v \mapsto \Lambda_{\mu_{f}}^{\prime}( v )$ around $v = s$ that
\begin{align}
\frac{ \alpha_{n} }{ n }
=
\Lambda_{\mu_{f}}^{\prime}( s_{n} )
=
a + \Lambda_{\mu_{f}}\mydprime( s ) \, (s_{n} - s) + \mathrm{O}( (s_{n} - s)^{2} )
\end{align}
uniformly on $\mathcal{I}_{n}$ as $n \to \infty$, which is equivalent to
\begin{align}
s_{n}
& =
s + \frac{ a_{n}( t ) }{ n \, \Lambda_{\mu_{f}}\mydprime( s ) } + \mathrm{O}( (s_{n} - s)^{2} )
\label{eq:Taylor0}
\end{align}
uniformly on $\mathcal{I}_{n}$ as $n \to \infty$.
Therefore, since $s_{n} \to s$ as $n \to \infty$ uniformly in $t \in \mathcal{I}_{n}$, we observe that
\begin{align}
s_{n} - s
=
\left( \frac{ a_{n}( t ) }{ n \, \Lambda_{\mu_{f}}\mydprime( s ) } \right) \, (1 + \mathrm{o}( 1 ))
\end{align}
uniformly on $\mathcal{I}_{n}$ as $n \to \infty$.
On the other hand, it follows from \eqref{eq:rate-function}, \eqref{eq:Taylor0}, and Taylor's theorem for $v \mapsto \Lambda_{\mu_{f}}( v )$ around $v = s$ that there exist real sequences $r_{f}^{(1)}(n, t) = \mathrm{o}( 1 )$ and $r_{f}^{(2)}(n, t) = \mathrm{O}( 1 )$ uniformly on $\mathcal{I}_{n}$ as $n \to \infty$ such that \eqref{eqn:last_block} (at the top of the next page) holds. 
\begin{figure*}
\begin{align}
\frac{ \alpha_{n} \, s_{n} }{ n } - \Lambda_{\mu_{f}}^{\ast}\left( \frac{ \alpha_{n} }{ n } \right)
& =
\Lambda_{\mu_{f}}( s_{n} )
\notag \\
& =
\Lambda_{\mu_{f}}( s ) + \Lambda_{\mu_{f}}^{\prime}( s ) \, (s_{n} - s) + \frac{ \Lambda_{\mu_{f}}\mydprime( s ) }{ 2 }  \, (s_{n} - s)^{2} + r_{f}^{(1)}(n, t) \, (s_{n} - s)^{3}
\notag \\
& =
a \, s - \Lambda_{\mu_{f}}^{\ast}( a ) + a \, (s_{n} - s) + \frac{ \Lambda_{\mu_{f}}\mydprime( s ) }{ 2 }  \, (s_{n} - s)^{2} + r_{f}^{(1)}(n, t) \, (s_{n} - s)^{3}
\notag \\
& =
\left( a - \frac{ \alpha_{n} }{ n } \right) s - \Lambda_{\mu_{f}}^{\ast}( a ) + \frac{ \alpha_{n} \, s_{n} }{ n } + \left( a - \frac{ \alpha_{n} }{ n } \right) \, (s_{n} - s) + \frac{ \Lambda_{\mu_{f}}\mydprime( s ) }{ 2 }  \, (s_{n} - s)^{2} + r_{f}^{(1)}(n, t) \, (s_{n} - s)^{3}
\notag \\
& =
\left( a - \frac{ \alpha_{n} }{ n } \right) s - \Lambda_{\mu_{f}}^{\ast}( a ) + \frac{ \alpha_{n} \, s_{n} }{ n } - \frac{ a_{n}( t )^{2} }{ n^{2} \, \Lambda_{\mu_{f}}\mydprime( s ) }
\notag \\
& \qquad  {}
+ \left( \frac{ a_{n}( t )^{2} }{ 2 \, n^{2} \, \Lambda_{\mu_{f}}\mydprime( s ) } \right) \, (1 + r_{f}^{(1)}(n, t))^{2} + r_{f}^{(2)}(n, t) \, \left( \frac{ a_{n}( t ) }{ n \, \Lambda_{\mu_{f}}\mydprime( s ) } \right)^{3} \, (1 + r_{f}^{(1)}(n, t))^{3}
\notag \\
& =
\left( a - \frac{ \alpha_{n} }{ n } \right) s - \Lambda_{\mu_{f}}^{\ast}( a ) + \frac{ \alpha_{n} \, s_{n} }{ n } - \frac{ a_{n}( t )^{2} }{ 2 \, n^{2} \, \Lambda_{\mu_{f}}\mydprime( s ) }
\notag \\
& \qquad  {}
+ \left( \frac{ a_{n}( t )^{2} }{ 2 \, n^{2} \, \Lambda_{\mu_{f}}\mydprime( s ) } \right) \, \Big( 2 \, r_{f}^{(1)}(n, t) + r_{f}^{(1)}(n, t)^{2} \Big) + r_{f}^{(2)}(n, t) \, \left( \frac{ a_{n}( t ) }{ n \, \Lambda_{\mu_{f}}\mydprime( s ) } \right)^{3} \, (1 + r_{f}^{(1)}(n, t))^{3} .\label{eqn:last_block}
\end{align}
\hrulefill
\end{figure*}
This is equivalent to
\begin{align}
\Lambda_{\mu_{f}}^{\ast}\left( \frac{ \alpha_{n} }{ n } \right)
& =
K_{f}(n, t) ,
\label{eq:Taylor1}
\end{align}
where $K_{f}(n, t)$ is defined as
\begin{align}
&K_{f}(n, s, t)\nonumber\\*
& \coloneqq n \, \Lambda_{\mu_{f}}^{\ast}( a ) + s \, a_{n}( t )  \nonumber\\* 
&\quad+ \frac{ a_{n}( t )^{2} }{ 2 \, n \, \Lambda_{\mu_{f}}\mydprime( s ) } \Bigg( 1 + r_{f}^{(1)}(n, t) \, \Big( 2 + r_{f}^{(1)}(n, t) \Big) \nonumber\\* 
&\qquad+ r_{f}^{(2)}(n, t) \, \Big( 1 + r_{f}^{(1)}(n, t) \Big)^{3} \, \frac{ 2 \, a_{n}( t ) }{ n \, \Lambda_{\mu_{f}}\mydprime( s )^{2} } \Bigg) .
\label{def:K_f}
\end{align}
Finally, it follows from Taylor's theorem for $v \mapsto \Lambda_{\mu_{f}}\mydprime( v )$ around $v = s$ that there exists a real sequence $r_{f}^{(3)}(n, t) = \mathrm{O}( 1 )$ uniformly on $\mathcal{I}_{n}$ as $n \to \infty$ such that
\begin{align}
\Lambda_{\mu_{f}}\mydprime( s_{n} )
& =
\Lambda_{\mu_{f}}\mydprime( s ) + \Lambda_{\mu_{f}}\mytrprime( s ) \, (s_{n} - s) + r_{f}^{(3)}(n, t) \, (s_{n} - s)^{2}
\notag \\
& =
\Lambda_{\mu_{f}}\mydprime( s ) \, (1 + R_{f}(n, t)) ,
\label{eq:Taylor2}
\end{align}
where $R_{f}(n, t)$ is defined as
\begin{align}
R_{f}(n, t)
&\coloneqq
\frac{ \Lambda_{\mu_{f}}\mytrprime( s ) \, (1 + r_{f}^{(1)}(n, t)) }{ \Lambda_{\mu_{f}}\mydprime( s ) } \, \frac{ a_{n}( t ) }{ n \, \Lambda_{\mu_{f}}\mydprime( s ) }\nonumber\\*
&\quad + \frac{ r_{f}^{(3)}(n, t) \, (1 + r_{f}^{(1)}(n, t))^{2} }{ \Lambda_{\mu_{f}}\mydprime( s ) } \, \left( \frac{ a_{n}( t ) }{ n \, \Lambda_{\mu_{f}}\mydprime( s ) } \right)^{2} .
\label{def:R_f}
\end{align}
By \eqref{eq:remainder_assumpsion}, we readily see that $R_{f}(n, t) = \mathrm{o}( 1 )$ uniformly on $\mathcal{I}_{n}$ as $n \to \infty$.
Applying the above asymptotic results to \lemref{lem:BR}, we obtain \lemref{lem:SLD}, as desired.
\hfill\IEEEQEDhere

\section*{Acknowledgement}
The authors are also grateful to the Associate Editor Prof.\ Tobias Koch and the  reviewers for their valuable and detailed comments.

\bibliographystyle{IEEEtran}
\bibliography{IEEEabrv,mybib}

\begin{IEEEbiographynophoto}{Yuta Sakai}(Member, IEEE)
was born in Japan in 1992.
He is currently an Assistant Professor at the Department of Electronics and Computer Science,
Graduate School of Engineering,
University of Hyogo. Her was formerly a Research Fellow in the Department of Electrical and Computer Engineering at the National University of Singapore (NUS) from 2018--2020.
He received the B.E.\ and M.E.\ degrees in the Department of Information Science from the University of Fukui in 2014 and 2016, respectively, and the Ph.D.\ degree in the Advanced Interdisciplinary Science and Technology from the University of Fukui in 2018.
His research interests include information theory and coding theory.
\end{IEEEbiographynophoto}

\begin{IEEEbiographynophoto}{Recep Can Yavas} (S'19) is currently a Ph.D.\ candidate in electrical engineering
at the California Institute of Technology (Caltech). He received the
B.S.\ degree from Bilkent University in Ankara, Turkey, in 2016 and the M.S.\
degree from Caltech in 2017, both in electrical engineering. His research
interests include information theory and probability theory.
\end{IEEEbiographynophoto}

\begin{IEEEbiographynophoto}{Vincent Y.\ F.\ Tan} (S'07-M'11-SM'15) was born in Singapore in 1981. He received the B.A.\ and M.Eng.\ degrees in electrical and information science from Cambridge University in 2005, and the Ph.D.\ degree in electrical engineering and computer science (EECS) from the Massachusetts Institute of Technology (MIT) in 2011. He is currently a Dean’s Chair Associate Professor with the Department of Electrical and Computer Engineering and the Department of Mathematics, National University of Singapore (NUS). His research interests include information theory, machine learning, and statistical signal processing.

Dr.\ Tan is a member of the IEEE Information Theory Society Board of Governors. He was an IEEE Information Theory Society Distinguished Lecturer from 2018 to 2019. He received the MIT EECS Jin-Au Kong Outstanding Doctoral Thesis Prize in 2011, the NUS Young Investigator Award in 2014, the Singapore National Research Foundation (NRF) Fellowship (Class of 2018), and the NUS Young Researcher Award in 2019. He is currently serving as an Associate Editor for the IEEE \textsc{Transactions on Signal Processing} and for the IEEE \textsc{Transactions on Information Theory}.
\end{IEEEbiographynophoto}

\end{document}